\documentclass[a4paper,11pt]{article}

\usepackage{geometry}										
\usepackage[english]{babel}									
\usepackage[utf8]{inputenc}									
\usepackage[T1]{fontenc}									

\usepackage{lmodern}										

\usepackage{amssymb,amsfonts,amsmath,amsthm}
\usepackage{booktabs}
\usepackage{hyphenat}
\usepackage{booktabs}
\usepackage{bbm}
\usepackage{multirow}
\usepackage{graphicx}
\usepackage{algorithm}
\usepackage{algorithmic}
\usepackage{bstnotations}

\newcommand{\nl}{\sigma_{\varepsilon}}
\newcommand{\ned}{N}
\newcommand{\xc}{\cx_\mathrm{c}}
\newcommand{\un}{U_\textrm{N}}
\newcommand{\gpn}{\sigma_{n}}
\usepackage{authblk}										

\title{Reliability analysis for data-driven noisy models using active learning}
\author[1]{Anderson V. Pires \thanks{apires@ethz.ch}}
\author[1]{M. Moustapha \thanks{moustapha@ibk.baug.ethz.ch}}
\author[1]{Stefano Marelli\thanks{marelli@ibk.baug.ethz.ch}}
\author[1]{Bruno Sudret\thanks{sudret@ethz.ch}}
\affil[1]{Chair of Risk, Safety and Uncertainty Quantification, ETH Z\"{u}rich, Switzerland}

\date{\today}

\usepackage{microtype}										

\usepackage{indentfirst}									
\usepackage{caption}
\usepackage{subcaption}

\usepackage{natbib}
\usepackage{hyperref}
\hypersetup{
pdftitle = {Reliability analysis for data-driven noisy models using active learning},
pdfauthor = {Anderson V. Pires, Maliki Moustapha,  Stefano Marelli, Bruno Sudret},
pdfsubject = {},
pdfkeywords = {Structural reliability, Noisy limit-state functions, Surrogate models, Gaussian process regression, Active learning},
colorlinks = false,                                                                                                                                                                          
}

\usepackage[capitalize]{cleveref}								
\creflabelformat{equation}{#2(#1)#3}								
\crefrangelabelformat{equation}{#3(#1)#4 to #5(#2)#6}						
\labelcrefformat{subequation}{#2(#1)#3}								
\labelcrefrangeformat{subequation}{#3(#1)#4 to #5(#2)#6}					

\graphicspath{{./Figures}} 

\begin{document}

\maketitle

\begin{abstract}
Reliability analysis aims at estimating the failure probability of an engineering system. It often requires multiple runs of a limit-state function, which usually relies on computationally intensive simulations. Traditionally, these simulations have been considered deterministic, \emph{i.e.,} running them multiple times for a given set of input parameters always produces the same output. However, this assumption does not always hold, as many studies in the literature report non-deterministic computational simulations (also known as noisy models). In such cases, running the simulations multiple times with the same input will result in different outputs. Similarly, data-driven models that rely on real-world data may also be affected by noise. This characteristic poses a challenge when performing reliability analysis, as many classical methods, such as FORM and SORM, are tailored to deterministic models.
To bridge this gap, this paper provides a novel methodology to perform reliability analysis on models contaminated by noise. In such cases, noise introduces latent uncertainty into the reliability estimator, leading to an incorrect estimation of the real underlying reliability index, even when using Monte Carlo simulation. To overcome this challenge, we propose the use of denoising regression-based surrogate models within an active learning reliability analysis framework. Specifically, we combine Gaussian process regression with a noise-aware learning function to efficiently estimate the probability of failure of the underlying noise-free model. We showcase the effectiveness of this methodology on standard benchmark functions and a finite element model of a realistic structural frame. \\[1em]


\end{abstract}

\maketitle

\section{Introduction}

Computer simulations are virtual representations of systems and are ubiquitous in current engineering practice. These virtual prototypes give insights into the behavior of the system under different scenarios, reducing the need for costly experimental setups. Such scenarios are controlled by the so-called \textit{input parameters}, a set of variables defined by the user which describe the characteristics and physical properties of the simulated system and its operating conditions.

Because these virtual models enable the simulation of the system under different conditions, they play a crucial role in reliability analysis \citep{Melchers2018}. The latter aims at quantifying the probability that the uncertainties in the input parameters lead to a performance failure. To compute such a probability of failure $P_f$, we consider a probabilistic framework where the uncertain input parameters are represented by a random vector denoted by $\boldsymbol{X}$. Their associated uncertainty can be entirely characterized by the joint probability density function (PDF) $f_{\boldsymbol{X}}$, defined on the domain $\cd_{\boldsymbol{X}} \subset \Rr^{M}$.

The state of the system, \emph{i.e.}, failure or safe operation, is defined through the so-called \emph{limit-state function} $g: \boldsymbol{x} \in \cd_{\boldsymbol{X}} \rightarrow \mathbb{R}$, which maps the input domain $\cd_{\boldsymbol{X}}$ to the real line. This function is constructed by comparing the response of the computational model to one or more limit-states of the system under consideration. Conventionally, failure occurs when $g\prt{\boldsymbol{x}}\leqslant 0$. The \emph{failure domain} $\cd_f$ can be defined as the subset of the input domain on which the limit-state function returns a non-positive value, \emph{i.e.}, $\cd_f = \acc{\boldsymbol{x}: g\prt{\boldsymbol{x}} \leqslant 0}$. Additionally, the particular set of points $\boldsymbol{x} \in \cd_{\boldsymbol{X}}$ such that $g\prt{\boldsymbol{x}} = 0$ is commonly referred to as the \emph{limit-state surface}. 
Given this framework, it is possible to define the probability of failure as follows \citep{Ditlevsen1996,Lemaire2009,Melchers2018}:
\begin{equation}
    P_f = \Pp\prt{g\prt{\boldsymbol{X}} \leqslant 0} = \int_{\cd_f} f_{\boldsymbol{X}}\prt{\boldsymbol{x}} \di{\boldsymbol{x}}.
    \label{eq_probability_of_failure}
\end{equation}

The integration domain depends on the response of the limit-state function $g$ and implicitly assumes that only a single scalar value $y = g\prt{\boldsymbol{x}_0}$ can be computed for any given vector $\boldsymbol{x}_0$, \emph{i.e.}, this definition relies on deterministic limit-state functions. Nevertheless, not all computational models are deterministic. The so-called \textit{non-deterministic models} provide stochastic responses for a given set of inputs, a behavior that, in practice, is seen as noisy.

The inherent variability observed in measurements acquired during physical experiments is a classic example of noise. However, this is not the only case where noise can be found. \cite{Giunta1994,Narducci_1995,Papila_2000,Forrester2006} report the existence of numerical noise in computational fluid dynamics (CFD) simulations, whereas \cite{Forrester2006,Duddeck_2007,Zhu__crashworthiness_2013, Paz_2020,Ahmadisoleymani_2021} report a similar behavior in crashworthiness simulations. \cite{Biermann_2008,Abbiati_2022} comment on noise in models combining physical experiments with simulations, known as \emph{hybrid simulators} in civil engineering. Similar issues also arise for the class of so-called grey-box models \citep{Chinesta_2018}, which combine data-based models with computational simulations.
Noise can also arise from intrinsic properties of the model, for instance in wind turbine simulations. In those models, wind loads are represented by three-dimensional random fields that depend on macroscopic parameters, such as the mean wind speed at a reference altitude, the turbulence intensity, the wind shear exponent, the air density and the blade pitch angle \citep{Dimitrov_2014,Abdallah_2019}. Despite fixing the value of their input parameters, infinitely many wind fields can be generated, leading to infinitely many wind turbine responses, characterizing the model response as noisy. In fact, the attention towards this class of models has grown in the past few years, with an increasing body of literature on so-called \textit{stochastic simulators} and \textit{emulators} \citep{Ankenman2009,Moutoussamy2015,AzziIJUQ2019,Torossian2020,Zhu2020,Zhu_2021a,Zhu_2021b,Zhu_2023a,Zhu_2023b,LuethenCMAME2023}.

This paper aims to address reliability analysis for limit-state functions corrupted by noise. Although only recently considered in this field, noise has been previously accounted for explicitly in Bayesian optimization. In general, the methods proposed in the latter capitalize on \textit{surrogate models}. These are inexpensive-to-evaluate mathematical approximations trained on a limited set of full-scale simulations called \emph{experimental design} (ED) and can be split into two broad classes: \emph{interpolation-} and \emph{regression-} based methods. Interpolation methods are generally used when assuming that the training points are not affected by random noise. Consequently, they precisely interpolate through the points of the ED. In contrast, regression methods aim at minimizing a global loss function, usually the mean squared error between the model predictions and the values collected in the experimental design. Surrogate models are usually employed to replace expensive simulations, and in the reliability analysis field, interpolation-based ones have often been the preferred choice. For noise-corrupted limit-state functions, however, regression-based surrogate models bring another advantage, as they can denoise the computational models \citep{Torre2019a}.

Many of the methods proposed in the Bayesian optimization field employ Gaussian process regression (GPR) as a surrogate \citep{Rasmussen2006}, with an emphasis on its adaptive version \citep{Jones1998,Shahriari_2016}, which involves sequentially enriching the experimental design in view of improving the current optimum. As most methods rely on GPR, the main difference between them is the so-called \textit{infill criteria}, a function that defines where the experimental design is enriched at each iteration. \cite{Picheny_2013a,Zhan_2020} review the available methods and infill criteria for the optimization of noisy models, with a focus on homoskedastic noise, \emph{i.e.}, when the noise variance is identical for all outputs. \cite{Jalali_2017} focus their review on heteroskedastic noise. More recently, techniques employing multi-fidelity models have been proposed by \cite{Ficini_2021} and \cite{Pellegrini_2022}.

In the context of reliability analysis,  \cite{chevalier_thesis_2013} addresses a noisy reliability problem related to nuclear safety. In their case, one of the parameters is computed through Monte Carlo simulation (MCS), which yields a small simulation noise due to the finite sample size used in practice. Similarly to the optimization literature, they capitalize on Gaussian process regression to denoise the problem and sequentially enrich the experimental design using the \textit{Stepwise Uncertainty reduction} (SUR) method \citep{Vazquez2009,Bect2012,Chevalier2014}. More recently, \cite{van_den_Eijnden_2021} presented a reliability problem with a high target reliability index, the estimation of which depends on a noise-corrupted limit-state function. Similarly, they employ GPR and extend the classical U learning function \citep{Echard2011} to cope with noisy data and sequentially enrich the experimental design.

Although a small number of works addressing the reliability analysis of mildly noisy problems can be found in the literature, a formalization of this problem, a detailed study of its effects, and a formal and robust methodology for coping with the challenges it poses are still lacking. In this paper, we aim to fill this gap by showing the particularities of performing reliability analysis on noisy models. Specifically, we show that traditional simulation-based methods for performing reliability analysis tend to converge to a wrong probability of failure, herein referred to as noisy probability of failure $\tilde{P}_f$. To circumvent this issue and converge to the actual noise-free $P_f$, denoising the problem is needed. To do that, we capitalize on regression-based surrogate models.

The paper is organized as follows: In Section 2, the problem of reliability analysis for noisy limit-state functions is first formulated. Section 3 demonstrates how the use of regression-based denoising leads to the estimation of the correct probability of failure. Section 4 outlines an efficient adaptive approach aimed at enhancing the computational efficiency of the algorithm. Section 5 showcases the results of the proposed methodology on two common academic examples and a complex, realistic structural frame problem. Finally, Section 6 presents our final thoughts and conclusions.

\section{Problem statement}
\label{sec_motivation}

In the context of reliability analysis for noisy models, we consider an underlying noise-free limit-state function $g:\boldsymbol{x} \in \cd_{\boldsymbol{X}} \subset \mathbb{R}^M$, whose values are, however, not directly accessible. Instead, we can measure or compute the value $\tilde{g}\prt{\boldsymbol{x}, \omega}$, which is corrupted by some zero-mean noise $\varepsilon\prt{\boldsymbol{x}, \omega}$. We pose: 
\begin{equation}
\tilde{g}\prt{\boldsymbol{x}, \omega} = g\prt{\boldsymbol{x}} + \varepsilon\prt{\boldsymbol{x}, \omega},
\label{eq_noisy_limit_state_function}
\end{equation}
\noindent where we assume $\Espe{\omega}{\varepsilon{\prt{\boldsymbol{x}, \omega}}} = 0$ for any $\boldsymbol{x} \in \cd_{\boldsymbol{X}}$. Note that $\varepsilon$ may or may not depend on $\boldsymbol{x}$ in practical applications.

The notation $\tilde{g}\prt{\boldsymbol{x}, \omega}$ explicitly introduces the random outcome $\omega$ of a sample space $\Omega$ that represents the variability attributed to the noise. In other words, when querying $g\prt{\boldsymbol{x}}$ several times, we get \textit{different} real values $\tilde{g}\prt{\boldsymbol{x},\omega_1}$, $\tilde{g}\prt{\boldsymbol{x},\omega_2}$ and so forth. 

We are interested in the probability of failure of the underlying noise-free limit-state function (\eqrefe{eq_probability_of_failure}):
\begin{equation}
    P_f = \Pp\prt{g\prt{\boldsymbol{X}} \leqslant 0},
    \label{eq_noise_free_Pf_limit_state_function}
\end{equation}

\noindent which we further refer to as the \textit{noise-free} probability of failure. In contrast, the so-called \textit{noisy} probability of failure is defined from the noisy limit-state function $\tilde{g}$ as follows
\begin{equation}
\tilde{P}_f = \Pp \prt{\tilde{g}\prt{\ve{X}, \omega} \leqslant 0} = \Pp\prt{g\prt{\ve{X}}+ \varepsilon\prt{\ve{X}, \omega} \leqslant 0}.
\label{eq_noisy_Pf_limit_state_function}
\end{equation}

Despite the noise unbiasedness, a different probability of failure is computed when blindly considering the noisy model and, in general, $\tilde{P}_f \geqslant P_f$. This occurs because the noise term introduces more uncertainty to the problem. To better illustrate this point, let us consider a simple linear limit-state function, such as the well-known $R-S$ case. For this problem, the random input vector contains two Gaussian variables $\boldsymbol{X}= \acc{R, S}$, where $R \sim \cn\prt{\mu_R, \sigma^2_R}$ corresponds to the resistance of the system, and $S \sim \cn\prt{\mu_S, \sigma^2_S}$ corresponds to the demand. Assuming that the noise term is independent on $R$ and $S$ and $\varepsilon\sim \cn\prt{0, \nl^2}$, the noise-free and noise-corrupted limit-state functions read:
\begin{equation}
g\prt{\boldsymbol{X}} = R-S \textrm{ and }  \tilde{g}\prt{\boldsymbol{X},\omega} = R-S + \varepsilon\prt{\omega}.
\end{equation}
\noindent Then, the analytical probabilities of failure are equal to
\begin{equation}
    P_f = \Phi \prt{-\frac{\mu_R - \mu_S}{\sqrt{\sigma^2_R + \sigma^2_S}}},
    \label{eq_RS_noise_free_solution}
\end{equation}
\noindent and
\begin{equation}
    \tilde{P}_f = \Phi \prt{-\frac{\mu_R - \mu_S}{\sqrt{\sigma^2_R + \sigma^2_S + \sigma^2_\varepsilon}}},
    \label{eq_RS_noisy_solution}
\end{equation}
\noindent respectively. From these two equations, it is clear that the only difference is the variance introduced by the noise term and that the noisy probability of failure $\tilde{P}_f$ is greater than the searched one $P_f$.

A crucial consequence of this observation is that simulation methods such as Monte Carlo simulation \citep{Rubinstein2016}, subset simulation \citep{Au2001} or importance sampling \citep{Melchers1989}  will converge to the noisy probability of failure (\eqrefe{eq_noisy_Pf_limit_state_function}). Similarly, approximation methods, such as the first and second-order reliability methods \citep{Hasofer1974,Rackwitz78}, will also fail in estimating the noise-free probability of failure, as these methods depend on the gradient of the limit-state function, which would be stochastic and lead to unstable results. In both cases, obtaining the noise-free probability of failure of the system (\eqrefe{eq_noise_free_Pf_limit_state_function}) is not possible.

\section{Retrieving the noise-free probability of failure}
Estimating the noise-free probability of failure while only having access to a noisy limit-state function is a matter of denoising. Following the approach from the works on optimization outlined earlier \citep{Picheny_2013a,Zhan_2020}, we propose the use of regression-based surrogate models to simultaneously denoise and cheaply estimate the probability of failure, as follows:
\begin{equation}
\hat{P}_{f}=\int_{\acc{\boldsymbol{x}: \hat{g}\prt{\boldsymbol{x}} \leqslant 0}} f_{\boldsymbol{X}}\prt{\boldsymbol{x}} \di \boldsymbol{x},
\label{eq_surrogate_Pf}
\end{equation}
\noindent where $\hat{P}_{f}$ and $\hat{g}$ respectively represent the estimated probability of failure and the surrogate of the limit-state function.

The regression component of the algorithm reconstructs a denoised version of the limit-state function, enabling the computation of the noise-free probability of failure. More precisely, because regression-based surrogate models are typically unbiased, they converge to the expected value of the noisy limit-state function (see, \eg \cite{Torre2019a,van_den_Eijnden_2021}). 

Examples of regression-based surrogate models include Gaussian process regression \citep{Rasmussen2006}, sparse polynomial chaos expansions \citep[PCE,][]{BlatmanJCP2011, Luethen_IJUQ_2022}, and support vector regression \citep[SVR,][]{Smola2004}. Due to its versatility and widespread adoption in the context of reliability analysis, in the following we focus on Gaussian process regression.

In principle, it is also possible to denoise the limit-state function by computing $\Espe{\omega}{\tilde{g}\prt{\boldsymbol{X}, \omega}}$ via direct Monte Carlo simulation, by simply drawing multiple realizations of the noisy limit-state function for each parameter set $\boldsymbol{x}$. However, this would require many replications of the Monte Carlo samples, which would in turn lead to an intractable computational costs, especially for small failure probabilities.


\subsection{Gaussian process regression basics}

Surrogate models are a widely-used approach to reduce the need for costly model evaluations. In this approach, the expensive model is considered as a \emph{black-box}, \emph{i.e.}, only its inputs and outputs are known. The aim is to learn the input-output mapping, using a limited-size experimental design $\ce$:
\begin{equation}
	\begin{aligned}
		\ce = \left\{ \prt{\boldsymbol{x}^{\prt{i}}, {y}^{\prt{i}}}\right.: y^{\prt{i}}  = \tilde{g} \prt{\boldsymbol{x}^{\prt{i}}, \omega^{\prt{i}}} \in \mathbb{R}, \left. \boldsymbol{x}^{\prt{i}} \in \cd_{\boldsymbol{X}} \subset \mathbb{R}^M, i = 1, \ldots, N   \right\},
	\end{aligned}
	\label{eq_ED_definition}
\end{equation}
where $\boldsymbol{x}^{\prt{i}}$ represents an $M$-dimensional vector of input parameters, $y^{\prt{i}}$ is the corresponding output from the noisy limit-state function and $N$ is the total number of available observations.

Gaussian process modeling \citep{Rasmussen2006} is a statistical technique frequently utilized in surrogate modeling. It assumes that the underlying function to approximate is one realization of an unknown Gaussian process. This model is generally expressed as:
\begin{equation}
\mathcal{M}^{\textrm{GP}}\prt{\boldsymbol{x}} = \mu\prt{\boldsymbol{x}} + \sigma^2 Z(\boldsymbol{x};\omega),
\end{equation}
where $\mu\prt{\boldsymbol{x}}$ is a trend, which is assumed constant and equal to zero herein (simple Kriging), $\sigma^2$ is the process variance and $Z(\boldsymbol{x};\omega)$ is a zero-mean, unit-variance stationary process fully characterized by its auto-correlation $R\prt{\boldsymbol{x},\boldsymbol{x}^\prime;\boldsymbol{\theta}}$ with hyperparameters $\boldsymbol{\theta}$.

While this model is aimed at approximating the underlying limit-state function $g\prt{\boldsymbol{x}}$, the observations in the experimental design are noisy. Gaussian process regression assumes an additive noise $\varepsilon$ which follows a zero-mean Gaussian distribution. We further assume that the noise is homoskedastic, \emph{i.e.}, it is the same for all observations. In this case, the covariance of the noise term is $\boldsymbol{\Sigma}_n^2 = \gpn^2 \, \ve{I}$, where $\gpn^2$ is the noise variance and $\ve{I}$ is the identity matrix of size $N \times N$.

By definition, the joint distribution between the response at a new location and the noisy observations $\mathcal{Y} = \acc{y^{\prt{i}}, i = 1, \ldots, N}$ is Gaussian and reads:
\begin{equation}
	\begin{Bmatrix}
		\widehat{g}(\boldsymbol{x}) \\
		\mathcal{Y} 
	\end{Bmatrix}
	\sim
	\mathcal{N}_{N+1}
	\begin{pmatrix}
		\ve{0}
		,
		& 
		
		\begin{Bmatrix}
			\sigma^2 & \ve{r}^T(\boldsymbol{x})  \\
			\ve{r}(\boldsymbol{x}) & \sigma^2\ve{R} +   \sigma^2_n \ve{I}
		\end{Bmatrix}
	\end{pmatrix},
\end{equation}
where $\ve{R}$ is an $N \times N$ auto-correlation matrix whose elements are defined by $R_{ij} = R\prt{\boldsymbol{x}^{(i)},\boldsymbol{x}^{(j)};\boldsymbol{\theta}}$ and $r\prt{\boldsymbol{x}}$ is the vector of cross-correlations whose elements are $\acc{r_i = R\prt{\boldsymbol{x},\boldsymbol{x}^{(i)}; \boldsymbol{\theta}}, \, i = 1, \ldots, N}$. 

By deriving the conditional distribution $\widehat{g}(\boldsymbol{x})|\mathcal{Y} \sim \mathcal{N}\prt{\mu_{\widehat{g}}(\boldsymbol{x}), \sigma_{\widehat{g}}^2(\boldsymbol{x})}$, we get the predictive equations for Gaussian process regression. Upon introducing the total process variance
\begin{equation}
	\sigma^2_{\textrm{total}} = \sigma^2 + \sigma^2_n,
\end{equation}
and the ratio
\begin{equation}
	\tau = \frac{\gpn^2}{\sigma_{\textrm{total}}^2},
\end{equation}
the mean and variance of $\widehat{g}\prt{\boldsymbol{x}}$ can be written as
\begin{equation}
	\mu_{\hat{g}}(\boldsymbol{x})= \tilde{\ve{r}}(\boldsymbol{x})^T \tilde{\ve{R}}^{-1}\cy,
	\label{eq_kriging_mean}
\end{equation}
\begin{equation}
	\sigma_{\hat{g}}^2(\boldsymbol{x})=\sigma_{\text {total }}^2\left(1-\tilde{\ve{r}}^T(\boldsymbol{x}) \tilde{\ve{R}}^{-1} \tilde{\ve{r}}(\boldsymbol{x})\right).
	\label{eq_kriging_variance}
\end{equation}
where $\tilde{\ve{r}} = \prt{1-\tau}\ve{r}$ and $\tilde{\ve{R}} = \prt{1-\tau}\ve{R} + \tau \, \ve{I}$ have been introduced for simplicity.

Eq.~\eqref{eq_kriging_mean} provides the Gaussian process prediction for a new location $\boldsymbol{x}$, while Eq.~\eqref{eq_kriging_variance} is the associated built-in error estimate. To fully define both equations, an auto-correlation function needs to be selected and its parameters, together with the process and noise variances, estimated. We consider here the Mat\'ern $5/2$ auto-correlation function, which reads:
\begin{equation}
	R\left(x, {x}^{\prime} ; \theta, \nu=5/2 \right) \left(1+\sqrt{5} \frac{\left|x-x^{\prime}\right|}{\theta}+\frac{5}{3}\left(\frac{\left|x-x^{\prime}\right|}{\theta}\right)^2\right) \exp \left[-\sqrt{5} \frac{\left|x-x^{\prime}\right|}{\theta}\right],
\end{equation}
The multi-dimensional case is derived using the following ellipsoidal formulation \citep{Rasmussen2006}:
\begin{equation}
	R\left(\boldsymbol{x}, \boldsymbol{x}^{\prime} ; \boldsymbol{\theta}\right)=R(h), \qquad h=\left[\sum_{i=1}^M\left(\frac{x_i-x_i^{\prime}}{\theta_i}\right)^2\right]^{\frac{1}{2}}.
\end{equation}

An analytical estimate of the total variance $\sigma_{\textrm{total}}^2$, which is a function of the auto-correlation hyperparameters $\boldsymbol{\theta}$, is derived by maximum likelihood \citep{DubourgThesis,Santner2003}:
\begin{equation}
\hat{\sigma}_{\textrm{total}}^2=\sigma_{\textrm{total}}^2(\boldsymbol{\theta})=\frac{1}{N}\mathcal{Y}^T \tilde{\boldsymbol{R}}^{-1}\mathcal{Y}.
\end{equation}

The hyperparameters $\boldsymbol{\theta}$ and the noise variance ratio $\tau$ are, in turn, jointly obtained by solving the following optimization problem:
\begin{equation}
	\hat{\boldsymbol{\theta}}, \hat{\tau}=\underset{\boldsymbol{\theta} \in \mathcal{D}_{\boldsymbol{\theta}}, \tau \in(0,1)}{\arg \min } \frac{1}{2}\left[\log (\operatorname{det} \tilde{\boldsymbol{R}})+ N \log \left(2 \pi \hat{\sigma}_{{\textrm{total} }}^2\right)+ N \right].
\end{equation}
In this paper, we use the implementation of the Gaussian process module of the \textsc{UQLab} software \citep{MarelliUQLab2014,UQdoc_20_105}.

\subsection{Estimation of ${P}_f$ via regression-based surrogate models}
\label{sec_proof_of_concept}

Once the surrogate model is trained, its mean predictor $\mu_{\hat{g}}$ can be used as a denoised version of the limit-state function $\tilde{g}$. Because $\mu_{\hat{g}}$ is a deterministic function, any reliability method can be used for computing $\hat{P}_f$. To demonstrate the feasibility of the proposed strategy, we denoise the $R-S$ function described in the previous section. To numerically solve the problem, we consider that  $R \sim \cn \prt{5, 0.8^2}$, and $S \sim \cn \prt{2, 0.6^2}$. The noise term is parametrically defined as $\varepsilon \sim \cn \prt{0, \sigma^2_{\varepsilon}}$, which allows testing the robustness of the method as the noise level $\nl^2$ increases. For this particular example, we evaluate $\hat{P}_f$ using Monte Carlo simulation \citep{Rubinstein2016}.

\begin{figure}[H]
    \centering
    \includegraphics[width=\linewidth]{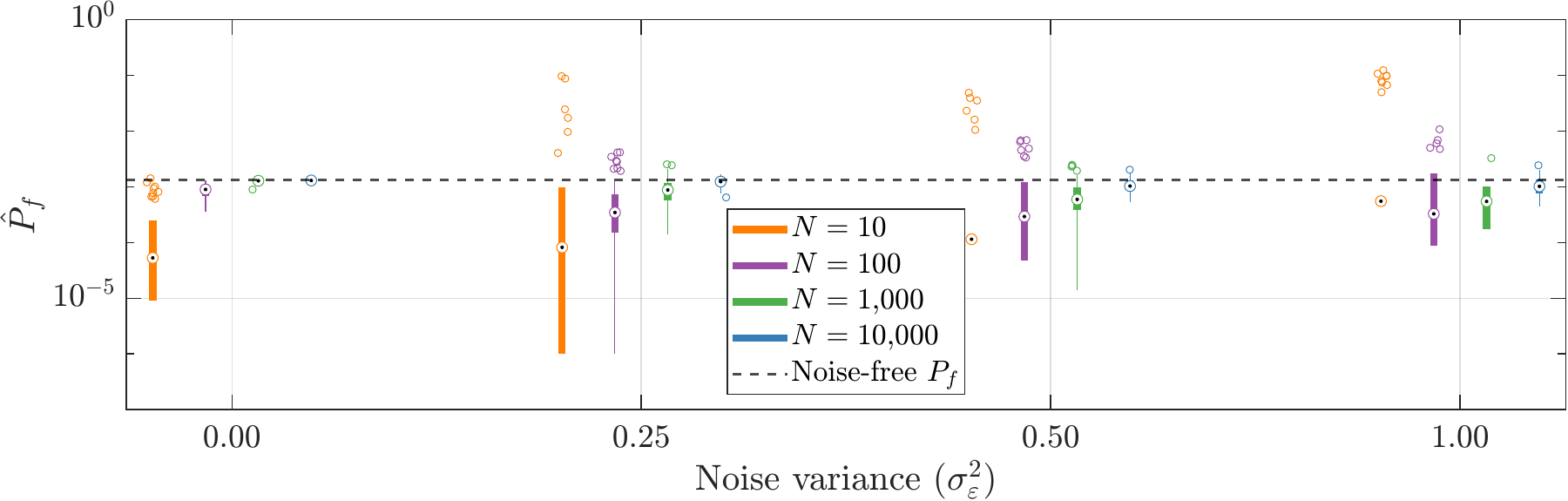}
    \caption{Noisy $R-S$ problem -- Estimation of the noise-free $P_f$ using a Gaussian process model based on $\ned$ points (obtained from the noisy limit-state function $\tilde{g}$ (\eqrefe{eq_noisy_limit_state_function})) and for different values of the noise variance.}
    \label{fig_denoising_largeED}
\end{figure}

\figref{fig_denoising_largeED} shows the probability of failure $\hat{P}_f$ computed for different noise levels $\nl^2 \in \acc{0; 0.25; 0.5; 1}$ and for different sizes $\ned$ of the experimental design used to fit the Gaussian process $\hat{g}\prt{r,s}$. The reference value, $P_f = \Phi\prt{-\frac{5-2}{\sqrt{0.8^2 + 0.6^2}}} = \Phi\prt{-3} = 1.35 \cdot 10^{-3}$, is shown with the horizontal dashed line. All results are represented by box plots since the whole procedure is replicated 50 times to account for statistical uncertainty.

For each level of noise, the noise-free probability of failure is recovered when $\ned$ is sufficiently large. The size $\ned$ required to get fairly accurate results increases with $\nl$, however; 100 points are sufficient when $\nl=0$ (noise-free limit-state function), whereas at least $1{,}000$ to $10{,}000$ are necessary for larger noise levels.

Nevertheless, relying on large EDs can be unfeasible in practice. First, obtaining such a large ED can be costly in the presence of computationally expensive limit-state functions. Second, as the size of the ED increases, the training and prediction steps can become expensive \citep{Chevalier2014}. To overcome these challenges and improve the general efficiency of the method, we employ an adaptive approach based on state-of-the-art active learning-based reliability analysis methods \citep{Teixeira2021, MoustaphaSS2022} developed for deterministic limit-state functions.

\section{Active learning methods for reliability analysis}

Active learning methods are iterative, surrogate-based approaches to efficiently solve complex reliability analysis problems \citep{Teixeira2021}. A Gaussian process is initially trained using an experimental design of small size, which is iteratively expanded. Essentially, the algorithm identifies at each iteration the most informative data points to add to the current ED, with the goal of maximizing the surrogate accuracy. Thanks to this rational selection of training points, this method significantly reduces the computational resources required for estimating the probability of failure.

\citet{MoustaphaSS2022} show that the vast majority of active learning methods feature four main components: a surrogate model, a reliability estimation method, a learning function and a stopping criterion. The surrogate model is the most crucial component of the method as its accuracy determines that of the estimated probability of failure. The reliability estimation method dictates how the probability of failure is calculated at each iteration. The learning function is responsible for selecting the point or set of points to be added to the ED at each iteration. The stopping criterion governs the termination of the algorithm. A step-by-step breakdown of the method and an illustration of how these components work together is shown in Algorithm~\ref{alg_actv_learn_method}.

\begin{algorithm}
    \caption{Pseudo-code for active learning reliability methods}
    \label{alg_actv_learn_method}
    \begin{algorithmic}
        \REQUIRE $\ce^{\prt{0}}$ (Initial ED)

        \STATE $i \leftarrow 0$

        \REPEAT

        \STATE Build \emph{surrogate model} $\hat{g}^{\prt{i}}\prt{\boldsymbol{x}; \ce^{\prt{i}}}$

        \STATE Compute $P_f^{\prt{i}}$ using $\hat{g}^{\prt{i}}\prt{\boldsymbol{x}; \ce^{\prt{i}}}$ and a \emph{reliability estimation method}

        \STATE Select $\boldsymbol{x}^{next}$ using the \emph{learning function}

        \STATE Enrich the ED: $\ce^{\prt{i+1}} \leftarrow \ce^{\prt{i}} \cup \acc{\boldsymbol{x}^{next} ,  \tilde{g}\prt{\boldsymbol{x}^{next}}} $

        \STATE $i \leftarrow i+1$

        \UNTIL{\emph{stopping criterion} is fulfilled}
    \end{algorithmic}
\end{algorithm}

We structure our active learning algorithm as follows. First, we select Gaussian process regression as surrogate model. In principle, any regression-based surrogate model that enables active learning, such as GPR, polynomial chaos expansions or support vector machines, can be used. For reliability estimation, we employ subset simulation, configured according to the \textit{overkill} setup introduced in \cite{MoustaphaSS2022}. Given that the enrichment step significantly impacts the efficiency and must take the noise into account in our case, a more comprehensive discussion is provided in the following section. Lastly, we establish a maximum budget of evaluations of the (noisy) limit-state function as the stopping criterion.

\subsection{Enrichment step}

During the enrichment step, we determine the next set of training points to include in the experimental design. This process is guided by a \emph{learning function} $\cl$, which is a utility function that assigns numerical scores to any point $\boldsymbol{x}$ within the domain. This score indirectly quantifies the information gained by introducing this point $\boldsymbol{x}$ as an additional training point. Notably, in the context of reliability analysis, these learning functions focus on minimizing the chance that the surrogate model makes a mistake in the sign of its prediction, leading to the misclassification of points on the safe domain as belonging to the failure domain, and vice-versa.

In theory, the scores provided by the learning function serve as the basis for solving an optimization problem that ultimately selects the points for inclusion in the ED. However, because computing the learning function is usually inexpensive, it is common practice to simplify this optimization problem into a discrete one, avoiding the need for complex optimization algorithms. In such cases, a \emph{candidate set} $\xc$ comprises a pre-defined collection of points that are potential candidates for inclusion in the ED. The learning function then computes the score of each point of $\xc$, which enables selecting the best enrichment point:
\begin{equation}
                  \boldsymbol{x}^{\text {next }}=\underset{\boldsymbol{x} \in \xc}{\arg \max } \, \cl(\boldsymbol{x}).
\end{equation}

\noindent Note that a batch of points can be selected at each step instead of a single one, using for instance clustering methods \citep{Dubourg2013, SchoebiASCE2017}.

Because Gaussian process models are the most commonly used surrogate models in reliability analysis, many learning functions capitalising on their properties are available in the literature, \eg \cite{Bichon2008,Echard2011,Lv_2015,Zhang_2019,Yi_2020, Shi_2020}. They are usually based on both the GP mean and variance. The critical distinction among these functions lies in how they define information. For instance, the $U$ learning function by \cite{Echard2011} identifies the point whose sign is most likely to be mistakenly predicted by the GP model, whereas the EFF \citep{Bichon2008} aims at estimating the potential improvement over the current prediction. In contrast, the $H$ learning function proposed by \cite{Lv_2015} relies on information entropy.

These learning functions generally aim to balance two competitive behaviors: \textit{exploration} and \textit{exploitation}. The former refers to the ability of the learning function to identify enrichment points in regions of high uncertainty and possibly discover multiple disjoint failure regions throughout the domain. The latter is related to the ability to efficiently refine the surrogate model in the vicinity of the current limit-state surface approximation. There are however two caveats to their use. On the one hand, learning functions with a highly explorative behavior may favor points in regions with insignificant probability mass and may lead to failure in accurately describing the limit-state surface. On the other hand, learning functions with highly exploitative behavior tend to select points close to one another, accurately describing a particular region of the limit-state surface but failing to find all failure regions.

All the learning functions mentioned above pose a challenge when dealing with noisy problems. They rely on the fact that the variance of the Gaussian process is zero at all points of the experimental design and tends to zero for points in their neighborhood. This is \textit{not} the case when considering Gaussian process regression for noisy limit-state functions. Indeed, the minimum GPR variance we can expect is the learned noise level $\gpn^2$. For this reason, these learning functions will show a strong tendency to repeatedly identify enrichment points within a confined region at the expense of other potential failure regions, a behavior reported by \cite{Wackers_2020,Ficini_2021}. To address this issue, noise-aware learning functions are needed. In the following section, we present a noise-aware extension of the well-known $U$ learning function from \cite{Echard2011}.

\subsubsection{Learning function $U_N$}
The $\un$ learning function, proposed by \cite{van_den_Eijnden_2021}, aims at identifying the point that will reduce the probability of misclassification $P_m$ the most. The latter was originally introduced by  \cite{Echard2011} and corresponds to the probability that the surrogate makes a mistake in the sign of a prediction, \emph{i.e.}, classifies as safe a sample in the failure domain or vice-versa. It can be computed as follows: 
\begin{equation} P_{m} \prt{\boldsymbol{x}} = \Phi\prt{-\frac{\abs{\mu_{\hat{g}}(\boldsymbol{x})}}{\sigma_{\hat{g}}(\boldsymbol{x})}}, \end{equation} 
 \noindent where $\mu_{\hat{g}}(\boldsymbol{x})$ is the surrogate prediction, $\sigma_{\hat{g}}(\boldsymbol{x})$ is the associated standard deviation of the GP, and $\Phi$ is the standard normal cumulative distribution function (CDF) \citep{Bect2012}.

The $\un$ learning function at a given point $\boldsymbol{x}$ is defined as the difference between the current probability of misclassification and the one computed after the point is added to the ED, that is:
 \begin{equation}
U_N\prt{\boldsymbol{x}} = P_{m} \prt{\boldsymbol{x}} - P_{m+1} \prt{\boldsymbol{x}},
\end{equation}
\noindent which is equivalent to,
\begin{equation}          
U_N\prt{\boldsymbol{x}} =\Phi\left(-\frac{\abs{\mu_{\hat{g}}(\boldsymbol{x})}}{\sigma_{\hat{g}}(\boldsymbol{x})}\right)-\Phi\left(-\frac{\abs{\mu_{\hat{g}+1}(\boldsymbol{x})}}{\sigma_{\hat{g}+1}(\boldsymbol{x})}\right),
\label{uq_Un_learning_function}
\end{equation}

\noindent where $P_{m+1} \prt{\boldsymbol{x}}$ is the residual probability of misclassification after $\boldsymbol{x}$ is added to the ED, while $\mu_{\hat{g}+1}(\boldsymbol{x})$ and $\sigma_{\hat{g}+1}(\boldsymbol{x})$ are the corresponding next step Gaussian process prediction and standard deviation.

\cite{van_den_Eijnden_2021} suggest estimating $P_{m+1} \prt{\boldsymbol{x}}$ through a one-step look-ahead approach. The major issue with this approach is that it requires computing the Gaussian process prediction after fitting the model with the additional training point for all candidate points. Because this would require running the expensive procedure for all $\boldsymbol{x} \in \xc$, the authors propose approximating $\mu_{\hat{g}+1}(\boldsymbol{x})$ by $\mu_{\hat{g}}(\boldsymbol{x})$. In this case, leveraging on the Gaussian properties of the model, it is possible to express the one-step look-ahead variance $\sigma^{2}_{\hat{g}+1}(\boldsymbol{x})$ and misclassification probability $P_{m+1} \prt{\boldsymbol{x}}$ with the following closed-form solutions:	
\begin{equation}
                  \sigma^{2}_{\hat{g}+1}(\boldsymbol{x})=\sigma^{2} _{\hat{g}}(\boldsymbol{x})\frac{\gpn^{2}}{\sigma^{2}_{\hat{g}}(\boldsymbol{x})+\gpn^{2}}
\label{eq_one_step_ahead_variance}
\end{equation}
and
\begin{equation}
    P_{m+1} \prt{\boldsymbol{x}} = \Phi\left(-\frac{\abs{\mu_{\hat{g}}(\boldsymbol{x})}}{\sigma_{\hat{g}+1}(\boldsymbol{x})}\right).
\end{equation}

\noindent As conclusion, the $\un$ learning function boils down to:
\begin{equation}
    \un\prt{\boldsymbol{x}} = \Phi\prt{-\frac{\abs{\mu_{\hat{g}}(\boldsymbol{x})}}{\sigma_{\hat{g}}(\boldsymbol{x})}} - \Phi\prt{-\frac{\abs{\mu_{\hat{g}}(\boldsymbol{x})} \sqrt{\sigma^{2}_{\hat{g}}(\boldsymbol{x})+\gpn^{2}}}{\sigma_{\hat{g}}(\boldsymbol{x}) \, \gpn }}.
\end{equation}

Finally, the next sample is chosen as follows:
\begin{equation}
                  \boldsymbol{x}^{\text {next }}=\underset{\boldsymbol{x} \in \xc}{\arg \max } \, U_{{N}}(\boldsymbol{x}).
                  \label{eq_optim_Un}
              \end{equation}

\subsubsection{Multi-point enrichment}
When multiple computing cores are available, enriching the surrogate models with batches of $K$ points at each iteration can significantly reduce the computation time of the reliability analysis \citep{SchoebiASCE2017}. However, as discussed in \cite{Chevalier2014}, this approach may result in a suboptimal experimental design compared to one of the same size enriched with one point at a time.

Many criteria for selecting multiple points at a time exist. \cite{Sacks1989} is one of the first works to address this issue and proposes a sequential look-ahead approach for selecting the multiple points. In this case, the learning function is run $K$ times for each step of the enrichment process, selecting a single point at each interaction. More recently, single-stage methods have been preferred. In this case, the $K$ points are found in the same loop. To achieve this, \cite{SchoebiASCE2017} suggests employing partitioning through weighted $K-$means, with the probability of misclassification serving as the weight. \cite{Lelivre_2018_AKMCSi} propose a similar approach but employ a different weighting method.

To alleviate the computational burden of clustering, we reduce the size of $\xc$ according to \cite{SchoebiASCE2017}. Then, we perform K-means clustering \citep{Zaki2014} on the reduced set of candidate points and select the optimal point within each given cluster. To perform the partition, the user must define the number $K$ of clusters in advance. We consider 1, 3 and 10 partitions in the sequel.


\section{Results on benchmark applications}
\subsection{Artificially corrupting noise-free models}
\label{sec_corrupting_methodology}

In this section, we show the results of our denoising methodology for two widely used benchmarks, the hat function and the four-branch series system, and for a finite element model of a realistic structural frame. Because these are noise-free models, we corrupt them with noise before testing the proposed methodology. In real applications with noisy models, this step would not be necessary.

All tested models are corrupted with additive noise, following \eqrefe{eq_noisy_limit_state_function}. Moreover, \cite{Picheny_2013a} suggest the noise to be normally distributed as this is widely accepted due to its simplicity. Consistent with this approach and considering homoskedastic unbiased noise, each realization of the noise component is sampled from independent and identically distributed (i.i.d) normal random variables $\cn\prt{0, \sigma_{\varepsilon}}$. Consequently, determining the noise level $\sigma_{\varepsilon}$ suffices for corrupting the noise-free model.

Specifying a suitable value for $\sigma_{\varepsilon}$ can be challenging, especially when comparing different benchmark functions, where no physical meaning can be linked to the noise level. The reason behind this is that the gradient of the limit-state function significantly affects how the failure probability responds to the noise variance. In general, functions with steeper gradients close to the limit state surface are more robust to noise, whereas functions with flatter regions are more sensitive to it. As an illustration, we introduce a constant $\gamma > 0$ for controlling the norm of the gradient in the previously mentioned $R-S$ example, as follows:
\begin{equation}
g\prt{\boldsymbol{X}} = \gamma\prt{R-S}  \textrm{ and }  \tilde{g}\prt{\boldsymbol{X},\omega} = \gamma\prt{R-S}  + \varepsilon\prt{\omega}.
\end{equation}

The norm of the gradient of the limit-state function is constant and corresponds to $\lVert \nabla \tilde{g} \rVert = \gamma \sqrt{2}$. The analytical noise-free and noisy probabilities of failure respectively read:
\begin{equation}
    P_f = \Phi \prt{-\frac{\mu_R - \mu_S}{\sqrt{\sigma^2_R + \sigma^2_S}}}
    \textrm{ and }
    \tilde{P}_f = \Phi \prt{-\frac{\gamma\prt{\mu_R - \mu_S}}{\sqrt{\gamma^2\prt{\sigma^2_R + \sigma^2_S} + \sigma^2_\varepsilon}}}.
\label{eq_noisy_Pf_function_of_gamma}
\end{equation}
\eqrefe{eq_noisy_Pf_function_of_gamma} shows that $\gamma$ does not affect $P_f$, but has an influence on its noisy counterpart. Moreover, to understand the relation between the gradient and the noise component, an asymptotic study can be carried out. In this case, a steep gradient corresponds to $\gamma \rightarrow \infty$, whereas a flat gradient is given by $\gamma \rightarrow 0$. Taking the limit of $\tilde{P}_f$ for these cases leads to:
\begin{equation}
\lim_{\gamma \rightarrow \infty}  \tilde{P}_f = P_f
\textrm{ and }
\lim_{\gamma \rightarrow 0}  \tilde{P}_f = 0.5.
\label{eq_limits_Pf}
\end{equation}
\eqrefe{eq_limits_Pf} shows that noise does not significantly impact limit-state surfaces with steep gradients, and the estimated $\tilde{P}_f$ converges to the noise-free one. On the other hand, it reveals that noise severely affects a rather flat limit-state surface, ultimately leading to a constant 50\% probability of failure, which corresponds to the probability of failure obtained for the noise term alone.

To take into account the gradient around the limit-state surface when corrupting the models, we propose defining $\sigma_{\varepsilon}$ as a function of the variance around its vicinity. Consequently, the obtained noise level can be made comparable across different benchmark limit-state functions. To this aim, we first introduce a \textit{region of interest} (ROI) by defining margins of the limit-state function using quantiles: 
\begin{equation}
ROI(\alpha; \mathcal{Y})  = \acc{y \in \mathcal{Y}: \abs{y} < q_{\alpha}} ,
\end{equation}
where $q_{\alpha}$ is defined by:
\begin{equation}
q_{\alpha} = \operatorname{inf}\{q \in \mathbb{R}: \Prob{\abs{g(\ve{X})} < q} > \alpha\},
\end{equation}
and $\cy$ is defined by:
\begin{equation}
    \mathcal{Y} = \left\{g(\boldsymbol{x}^{(1)}), \ldots,  g(\boldsymbol{x}^{(n)}) \right\},
\end{equation}

\noindent where $\acc{\boldsymbol{x}^{(1)}, \ldots, \boldsymbol{x}^{(n)}}$ corresponds to samples drawn according to $f_{\ve{X}}$.

Finally, our proxy to the level of introduced noise is defined as the variance of the samples inside the ROI, as follows:
\begin{equation}
\sigma_{\varepsilon}^2(\alpha) \stackrel{\text{def}}{=}  \Var{g(\boldsymbol{X}) \bigm\vert g(\boldsymbol{X}) \in ROI}.
\label{eq_noise_level}
\end{equation}

With this approach, the noise level is a function of $\alpha \in \prt{0,1}$. The closer $\alpha$ is to 0, the milder the noise is, independently of the limit-state function.
As this method requires simulating many samples, it can only be used for inexpensive benchmark functions. In actual applications, however, a physical meaning should be attributed to the noise level whenever possible, typically related to the expected measurement or numerical noise levels. Again, the procedure described in this section is only used to build coherent noisy benchmark functions and facilitate cross-comparison.

\subsection{Four-branch function}

The four-branch function, initially proposed by \citet{Waarts2000}, has been commonly used to benchmark reliability estimation algorithms. The corresponding reliability analysis problem has a total of four distinct failure regions. Converging to the correct probability of failure is, therefore, non-trivial, and methods such as FORM/SORM and importance sampling tend to significantly underestimate the probability of failure. The noise-free limit-state function reads:
\begin{equation}
    g(\boldsymbol{x})=\min \left\{\begin{array}{c}
        3+0.1\left(x_{1}+x_{2}\right)^{2}-\frac{x_{1}+x_{2}}{\sqrt{2}} \\
        3+0.1\left(x_{1}+x_{2}\right)^{2}+\frac{x_{1}+x_{2}}{\sqrt{2}} \\
        \left(x_{1}-x_{2}\right)+\frac{6}{\sqrt{2}}                    \\
        \left(x_{2}-x_{1}\right)+\frac{6}{\sqrt{2}}
    \end{array}\right\}.
\end{equation}

This problem comprises two random input variables, modeled as independent standard normal distributions, \emph{i.e.}, $X_{1},X_{2} \sim \cn\prt{0,1}$. The reference noise-free probability of failure is computed via Monte Carlo simulation using $10^7$ samples and is equal to $4.51 \cdot 10^{-3}$. The associated coefficient of variation is smaller than 1\%.

\begin{figure}[H]
     \centering
     \begin{subfigure}[c]{0.49\textwidth}
         \centering
         \includegraphics[height = 7cm]{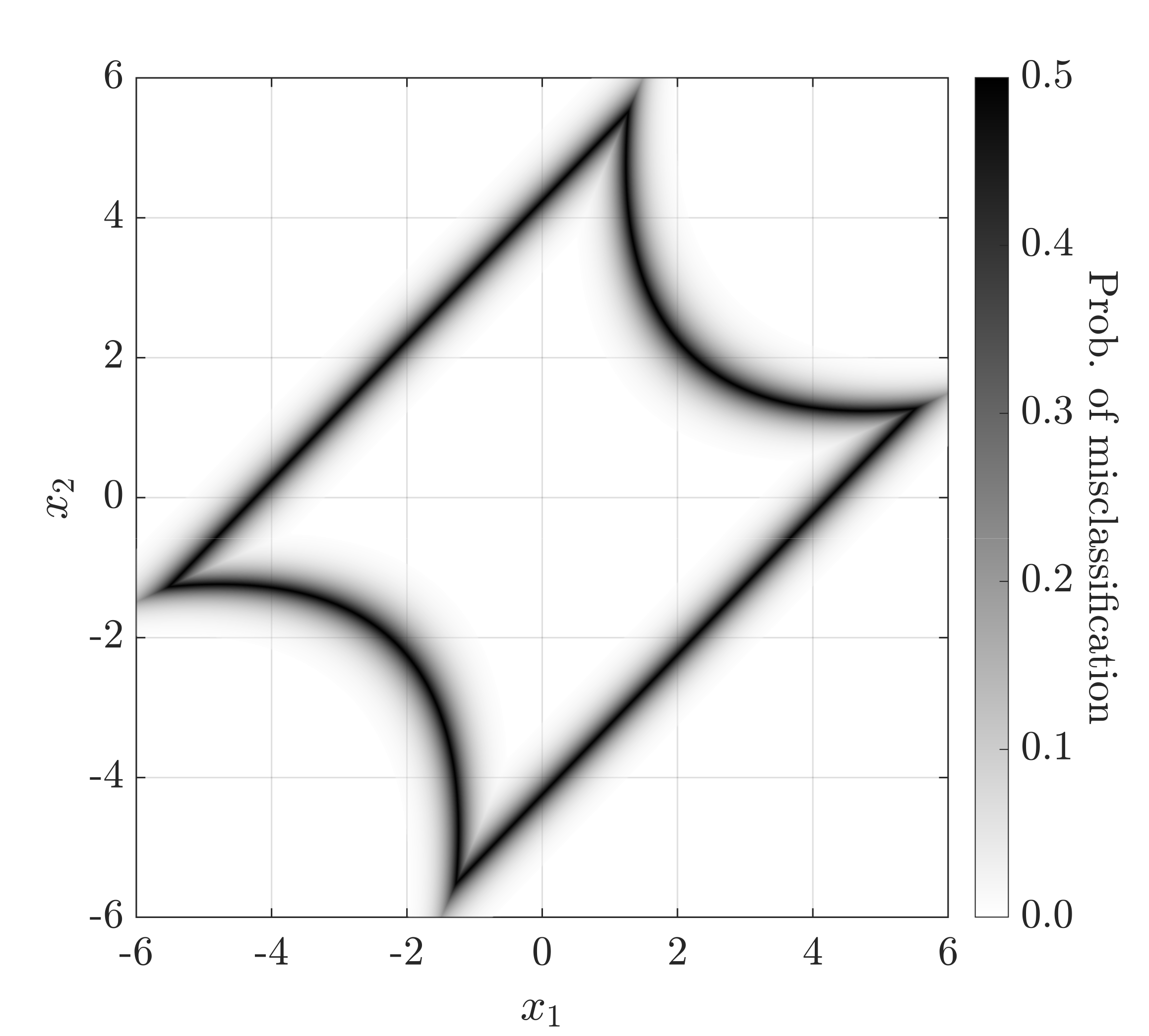}
         \caption{}
    \label{fig_4b_prob_of_misclassification}
     \end{subfigure}
     \begin{subfigure}[c]{0.49\textwidth}
         \centering
         \includegraphics[height = 7cm]{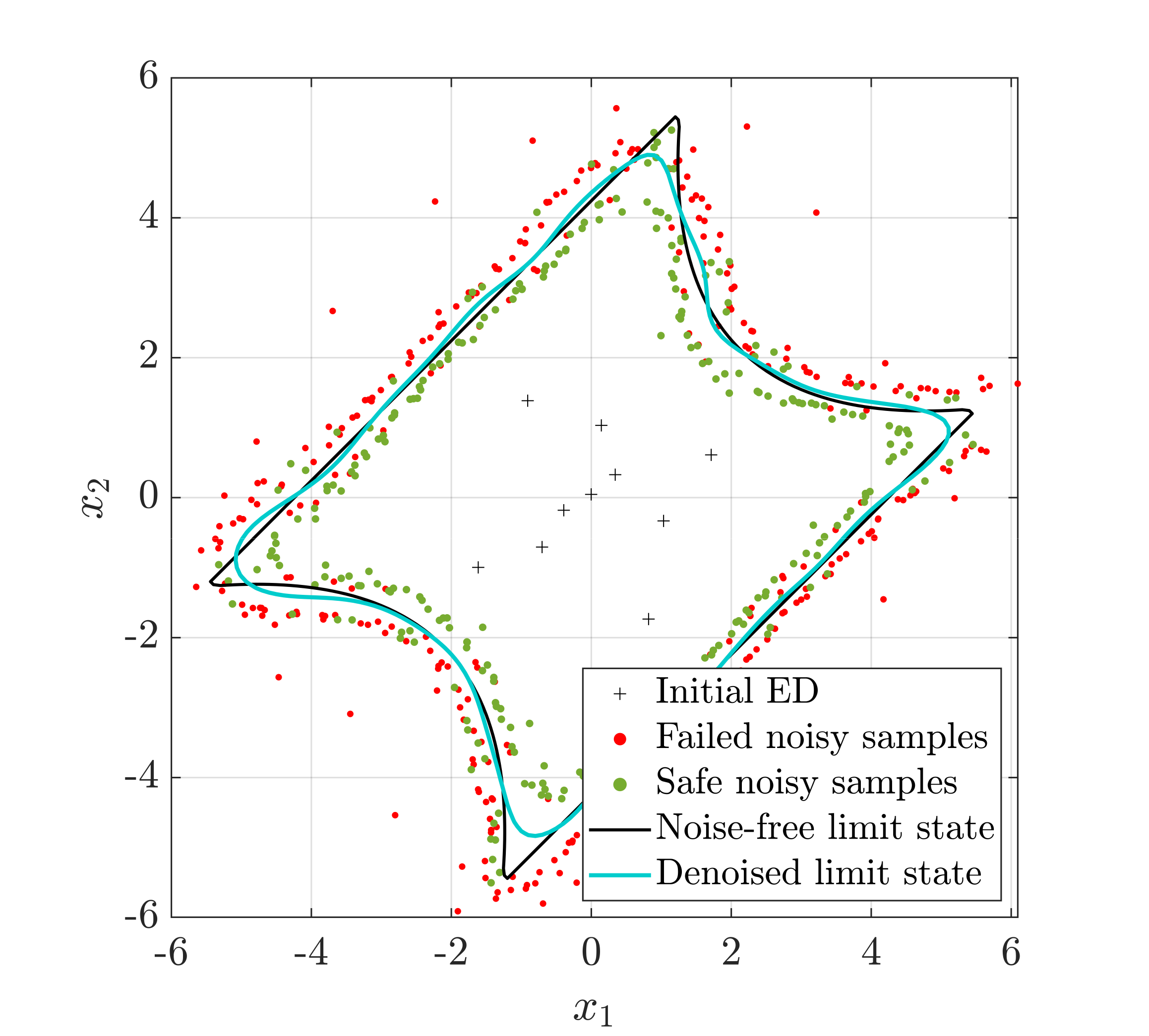}
         \caption{}
    \label{fig_4b_denoised_lmit_state_surface}
     \end{subfigure}
     \caption{Four-branch function -- (a): heatmap of the probability of misclassification as described in \eqrefe{eq_actual_misclassification_prob} for a noise level of $\alpha = 0.05$. (b): noise-free limit-state surface (black line), final limit-state surface estimated by the GPR (cyan line), initial ED (black cross). The green (resp. red) dots are the safe (resp. failed) samples selected by the $\un$ learning function for a noise level of $\alpha = 0.05$.}
\end{figure}

We have chosen this model to test our denoising methodology, because learning functions generally tend to get stuck in exploitative loops for this problem, \emph{i.e.}, they fail to discover the four branches. This toy example is corrupted according to the methodology described in Section \ref{sec_corrupting_methodology}. Due to noise corruption, depicting the limit-state surface is impossible. Instead, \figref{fig_4b_prob_of_misclassification} depicts a heatmap of the probability of misclassification caused solely by the noise component, defined as:
\begin{equation}
\tilde{P}_m\prt{\boldsymbol{x}} = \Phi \prt{-\frac{\abs{{g}\prt{\boldsymbol{x}}}}{\sigma_{\varepsilon}}}.
\label{eq_actual_misclassification_prob}
\end{equation}

\noindent \figref{fig_4b_prob_of_misclassification} was obtained for a model corrupted with $\alpha=0.05$.

The Gaussian process model is trained with an initial ED containing $N_{\textrm{ini}}= 10$ points, obtained via space-filling Latin hypercube sampling \citep[LHS,][]{McKay1979}. Enrichment is performed with the $U_N$ learning function, and three enrichment strategies are tested by adding 1, 3 or 10 points per iteration. Finally, the algorithm is terminated after 600 points are added. Each setup of the problem is repeated 50 times with different initial experimental designs.

\begin{figure}[H]
     \centering
     \begin{subfigure}[b]{0.48\textwidth}
         \centering
         \includegraphics[width=.9\textwidth]{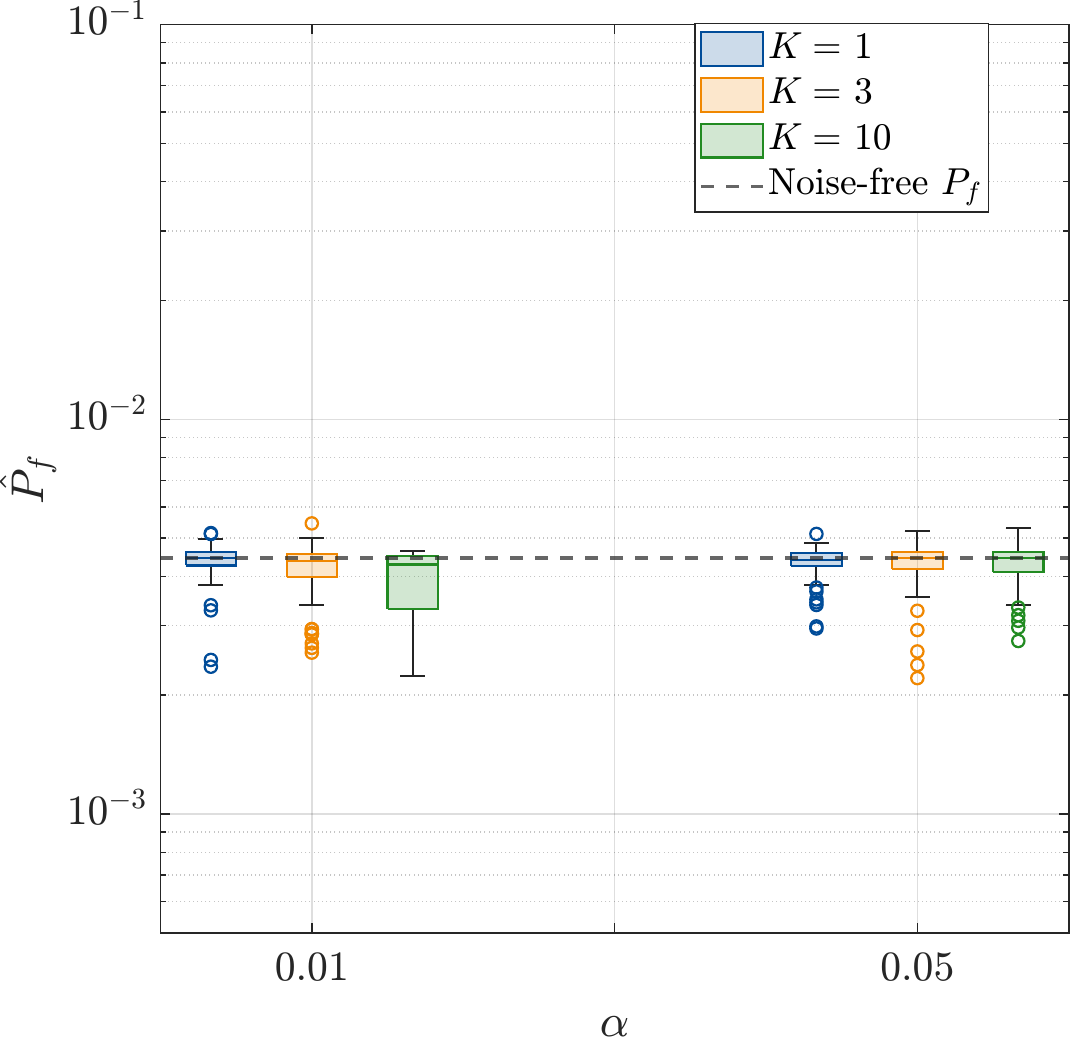}
        \caption{}
    \label{fig_4b_boxplots_Un}
     \end{subfigure}
     \hfill
     \begin{subfigure}[b]{0.48\textwidth}
         \centering
    \includegraphics[width=.9\textwidth]{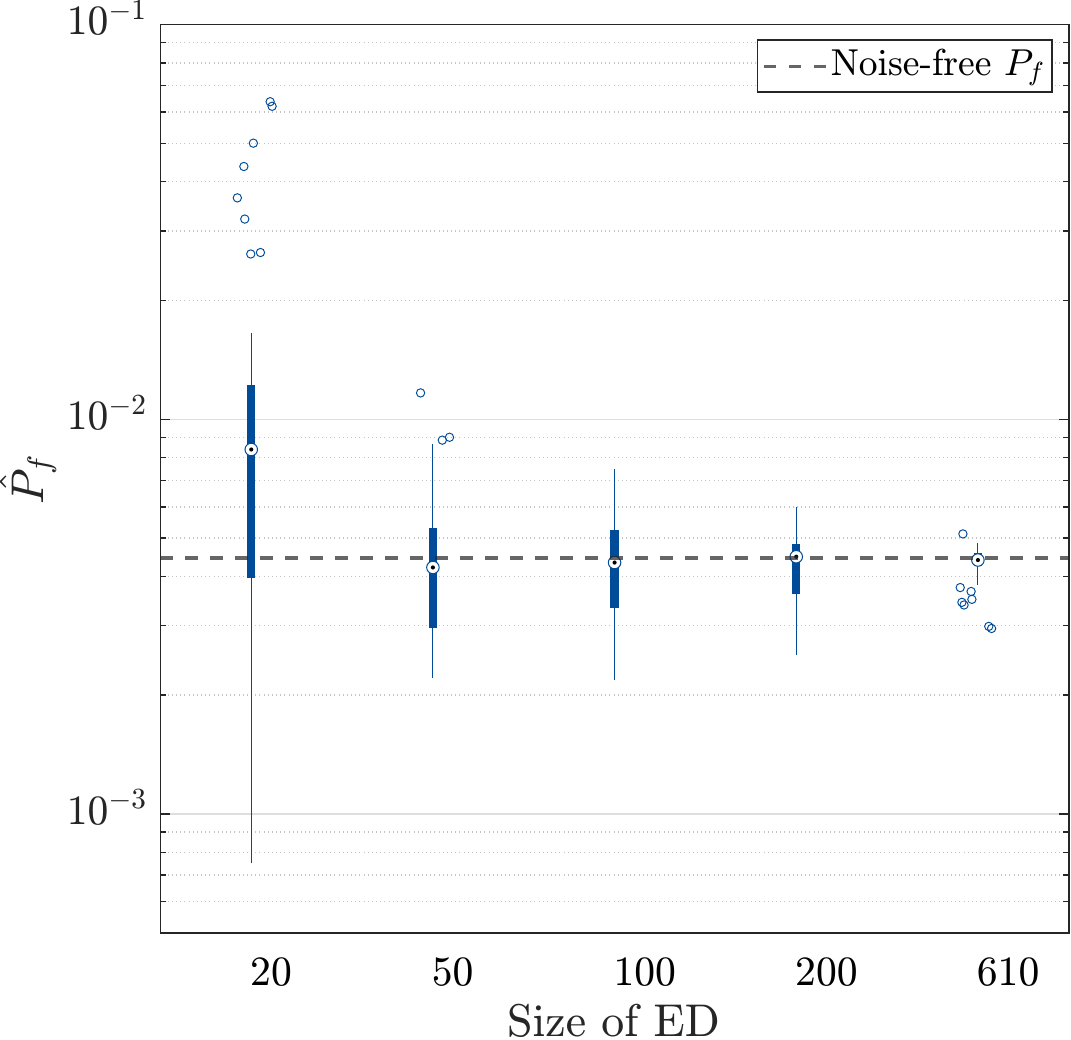}
    \caption{}
    \label{fig_4b_convergence_boxplots}
     \end{subfigure}
     \caption{Four-branch function -- (a): final probability of failure estimated using GPR with learning function $\un$. (b): boxplots depicting the convergence behavior for $\alpha = 0.05$ and $K=1$.}
     \label{fig_4b_boxplots}
\end{figure}

\figref{fig_4b_boxplots} depicts the results of these experiments for two different levels of noise: $\alpha = 0.01$ and $0.05$. \figref{fig_4b_boxplots_Un} shows that the learning function can converge to the noise-free $P_f$ in all tested scenarios. Moreover, no significant difference in performance is observed for multi-point enrichment. \figref{fig_4b_convergence_boxplots} depicts the boxplots showing the convergence behavior of the experiments. Additionally, it is possible to observe that the model converges to $P_f$ with around 200 points, suggesting that a proper stopping criterion might reduce the required computational effort. Finally, \figref{fig_4b_denoised_lmit_state_surface} depicts the denoised limit-state surface for the experiment that leads to the median $\hat{P}_f$ for $\alpha = 0.05$. The observed alignment of the points around the limit-state surface can be attributed to a limit behavior of the $\un$ learning function, a feature explored in greater detail in Appendix~\ref{appendix}.

\subsection{Hat function}

The Hat function is another frequently used benchmark in reliability analysis. It is a polynomial function that reads:
\begin{equation}
    g\prt{\boldsymbol{x}} = 12 - \prt{x_1 - x_2}^2 - 8\prt{x_1+x_2-4}^3.
    \label{eq_hat_noise_free}
\end{equation}
The random input vector comprises two i.i.d normal random variables $X_1, X_2 \sim \cn \prt{0.25, 1}$. The reference $P_f=9.76 \cdot 10^{-4}$ was computed via Monte Carlo simulation using $2\cdot 10^7$ samples, and its associated coefficient of variation is smaller than $1\%$.

\begin{figure}[H]
     \centering
     \begin{subfigure}[c]{0.49\textwidth}
         \centering
         \includegraphics[height = 7cm]{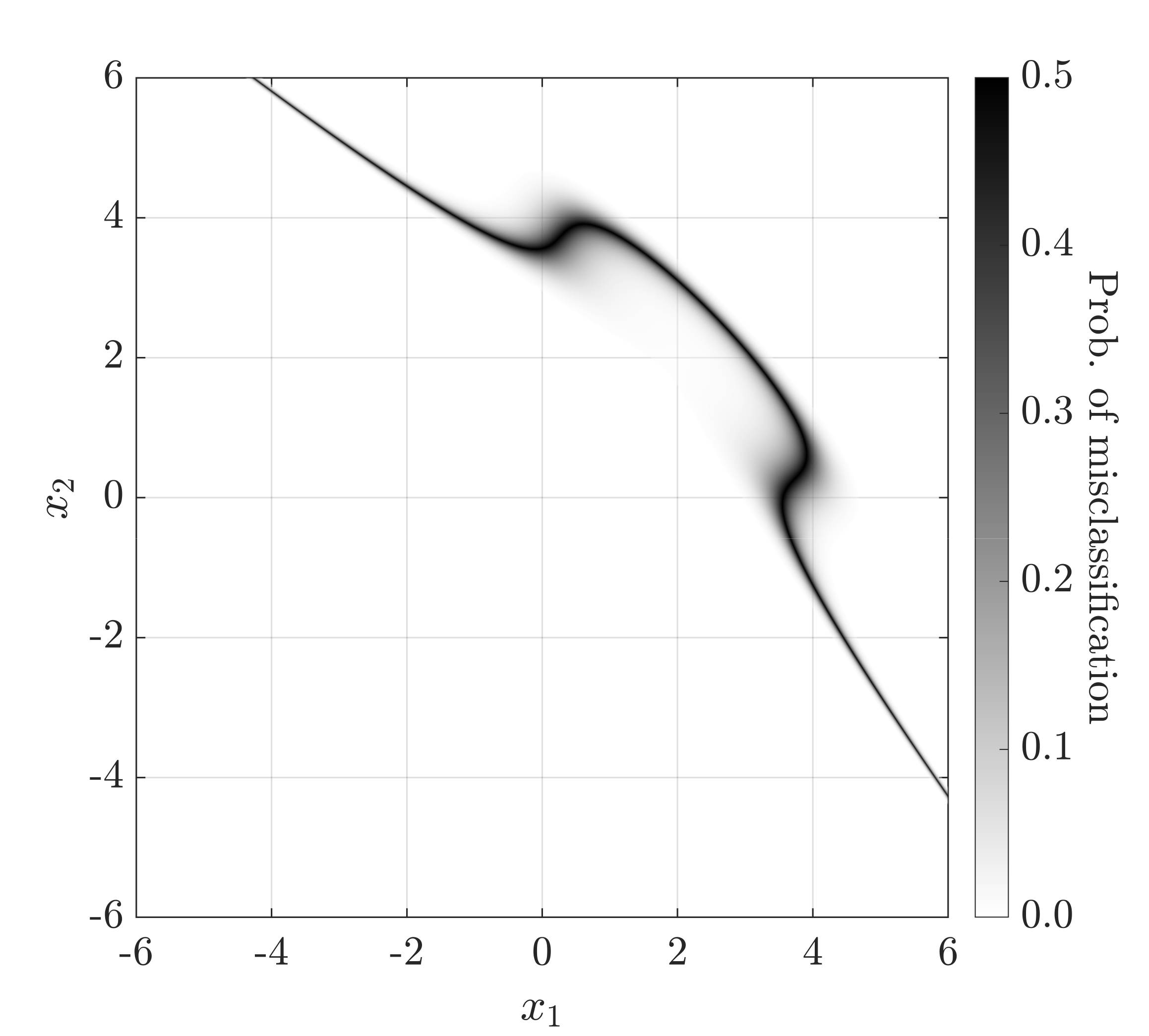}
         \caption{}
    \label{fig_hat_prob_of_misclassification}
     \end{subfigure}
     \begin{subfigure}[c]{0.49\textwidth}
         \centering
         \includegraphics[height = 7cm]{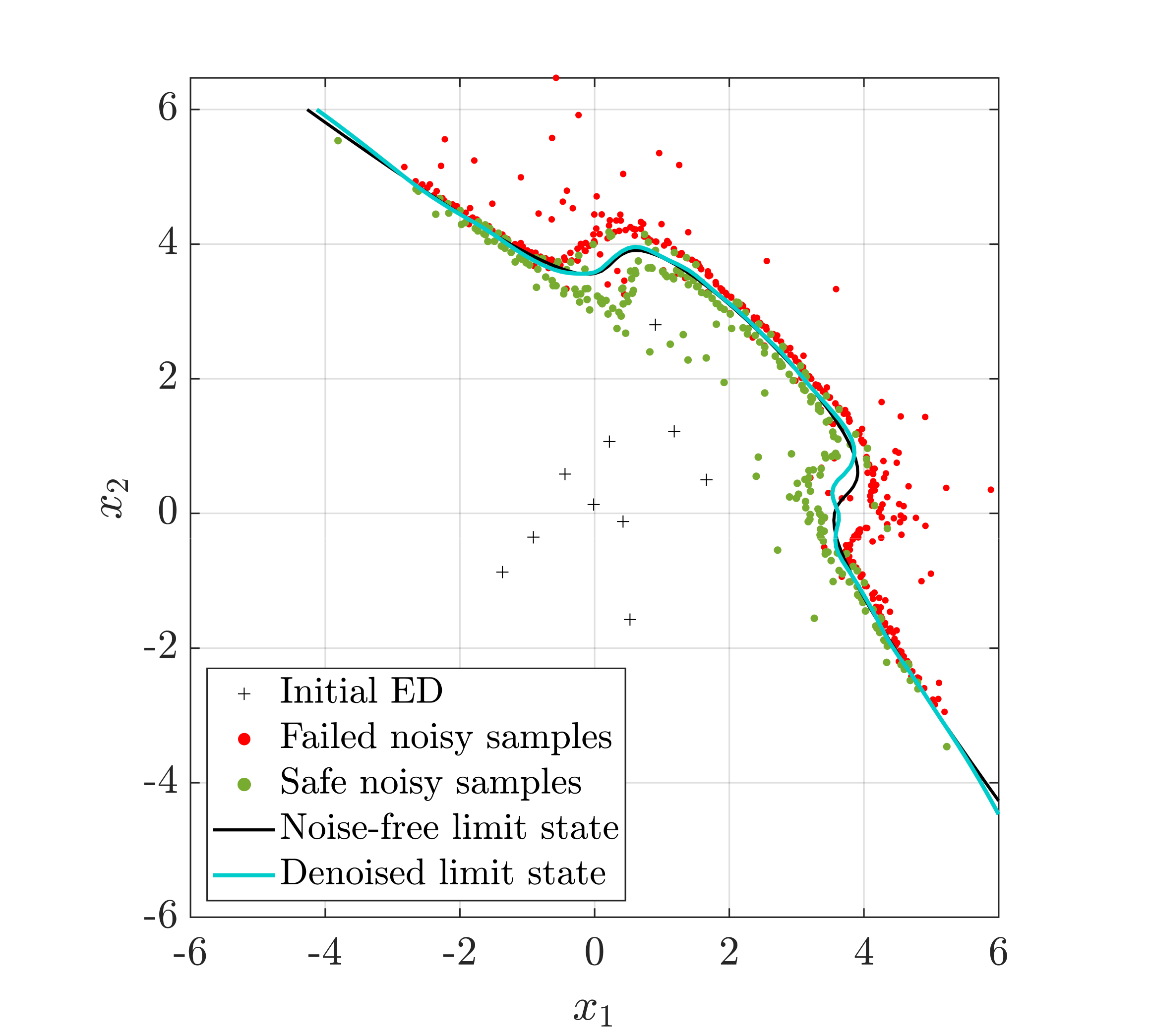}
         \caption{}
    \label{fig_hat_denoised_lmit_state_surface}
     \end{subfigure}
     \caption{Hat function -- (a): heatmap of the probability of misclassification as described in \eqrefe{eq_actual_misclassification_prob} for a noise level of $\alpha = 0.05$. (b): noise-free limit-state surface (black line), final limit-state surface estimated by the GPR (cyan line), initial ED (black cross). The green (resp. red) dots are the safe (resp. failed) samples selected by the learning function.}
\end{figure}

Similarly to the previous example, the limit-state function is corrupted according to the methodology described in Section \ref{sec_corrupting_methodology}. In this example, the limit-state surface lies in a flat region (small gradient), which hinders the denoising process, as discussed above. A graphical illustration of this issue is given in \figref{fig_hat_prob_of_misclassification}, which depicts the heatmap of the probability of misclassification for $\alpha =0.05$. A considerable area around the limit-state surface presents a significant probability of misclassification, since even a small amount of noise can swap the sign of $g\prt{\boldsymbol{x}}$.

Similarly to the previous example, 50 experiments are carried out, and the size of the initial ED for each replication corresponds to $N_{\textrm{ini}} =  10$ points obtained via LHS. Moreover,  three enrichment strategies are tested, adding 1, 3 and 10 points at a time. The algorithm is terminated after 600 points are added to the experimental design.

\begin{figure}[H]
     \centering
     \begin{subfigure}[t]{0.48\textwidth}
         \centering
\includegraphics[width=.9\textwidth]{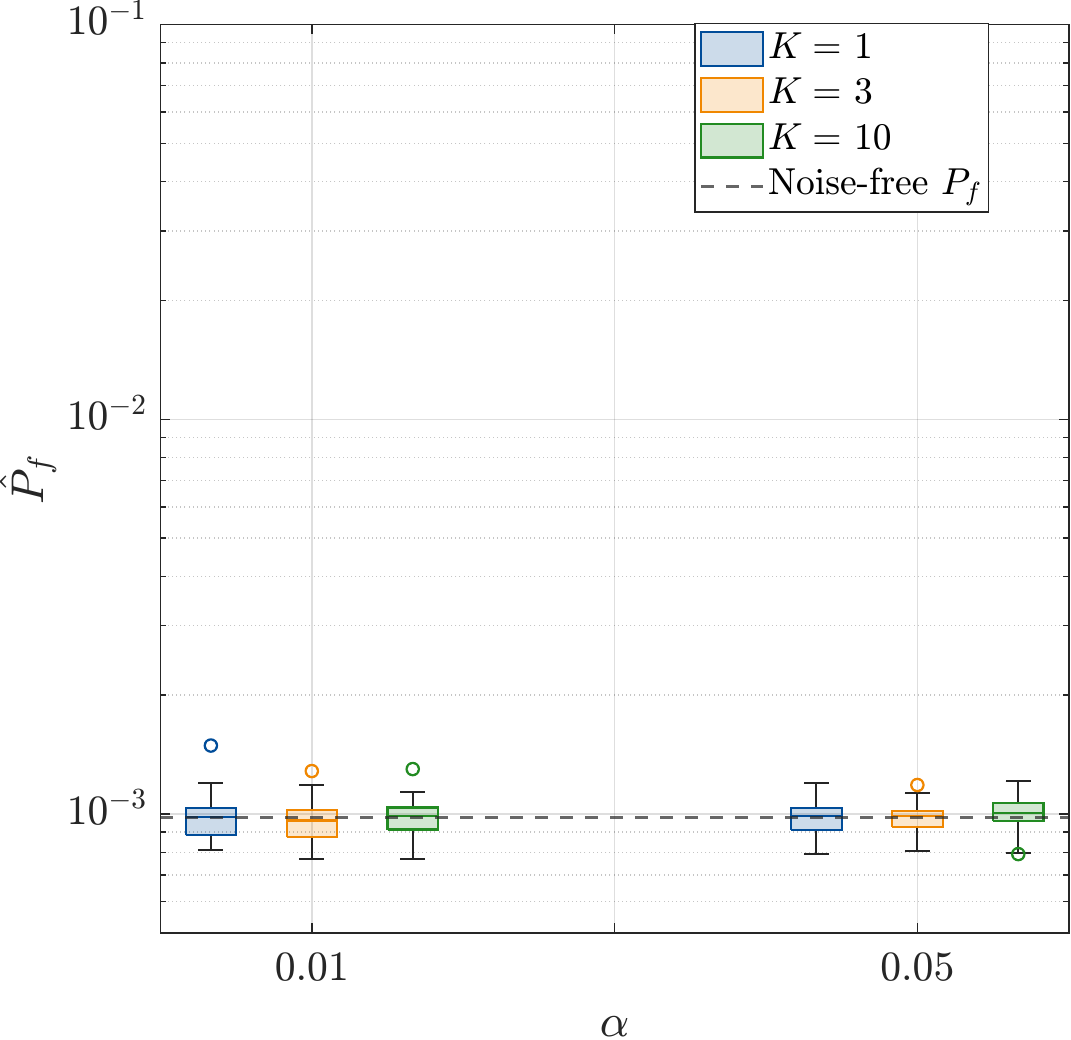}
         \caption{}
    \label{fig_hat_boxplots}
     \end{subfigure}
     \hfill
     \begin{subfigure}[t]{0.48\textwidth}
         \centering
         \includegraphics[width=.9\textwidth]{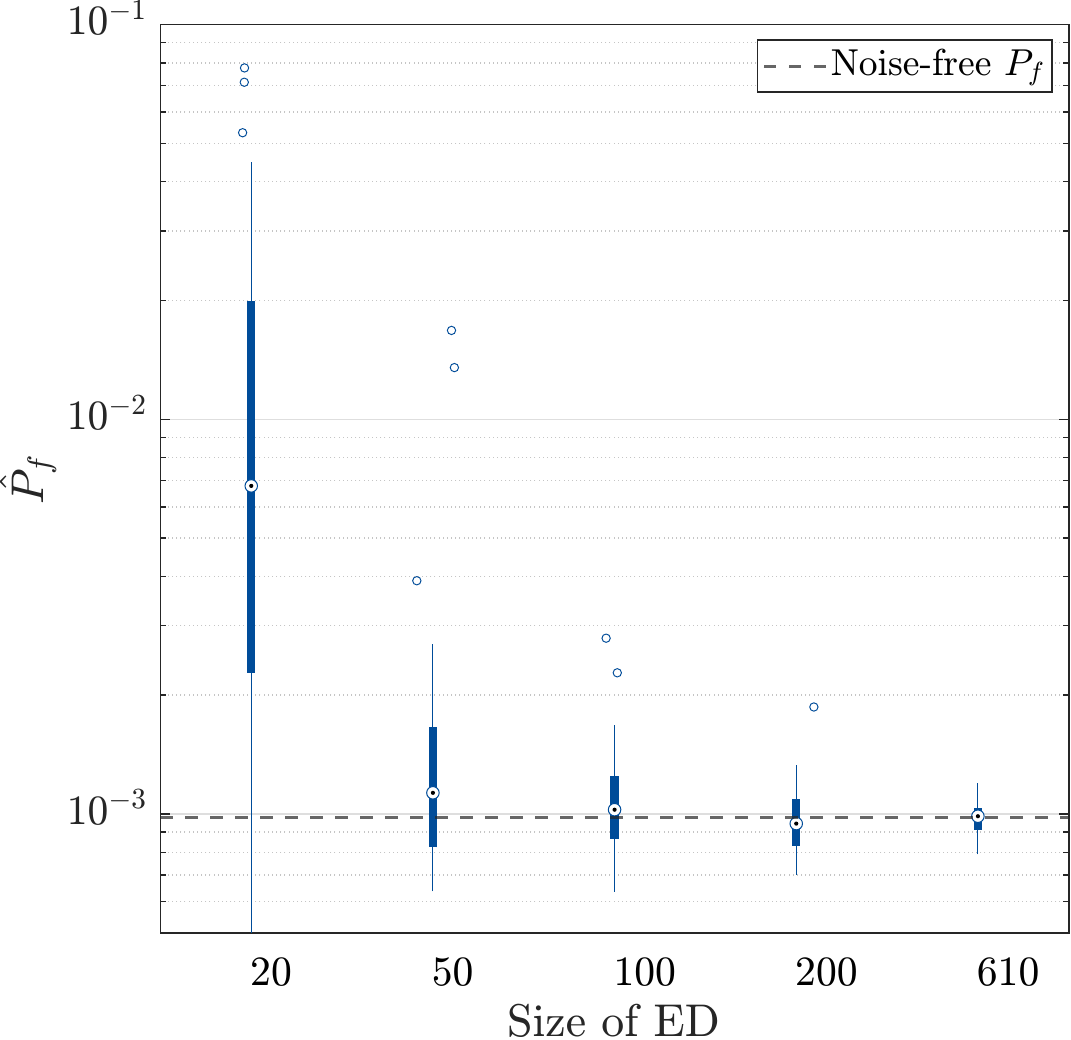}
         \caption{}
    \label{fig_hat_boxplot_convergence}
     \end{subfigure}
     \caption{Hat function -- (a): final probability of failure estimated using GPR with learning function $\un$. (b): boxplots depicting the convergence behavior for $\alpha = 0.05$ and $K=1$.}
     \label{fig_hat_results}
\end{figure}

\figref{fig_hat_boxplots} depicts the boxplots of the computed probability of failure for two levels of noise. All setups converge accurately to the noise-free probability of failure. \figref{fig_hat_boxplot_convergence} showcases the convergence of the estimator $\hat{P}_f$ with respect to the size of the experimental design. It reveals that the algorithm converged to the noise-free probability of failure already at around 200 samples. Finally, \figref{fig_hat_denoised_lmit_state_surface} depicts the denoised limit-state surface and final ED of the experiment, which yielded the median $\hat{P}_f$ for $\alpha = 0.05$. 

\subsection{Structural frame}

\begin{figure}[H]
          \centering
          \includegraphics[width=0.8\linewidth]{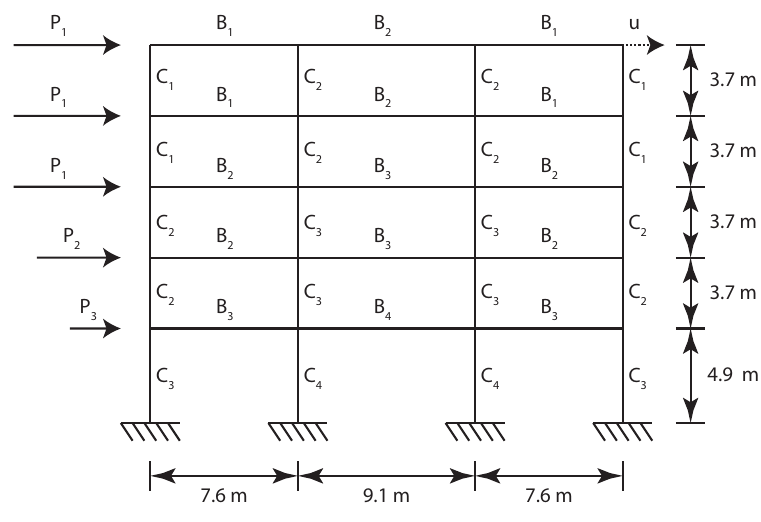}
          \caption{Sketch of a 3-span, 5-story structural frame.}
          \label{fig_SF_sketch}
        \end{figure}

Real-world applications are generally high-dimensional and more complex than the previous benchmark examples. To showcase the performance of the algorithm for such cases, we apply the methodology to the finite element model of a structural frame depicted in \figref{fig_SF_sketch}. This problem was first introduced in the reliability analysis literature by \cite{LiuPL91} and has often been used in the context of surrogate modeling and reliability analysis \citep{BlatmanPEM2010,MarelliSS2018,MoustaphaSS2022}. Of interest is the probability that the horizontal displacement $u$ of the right top corner of the building exceeds a critical threshold $u_{max}$. The associated limit-state function reads:
\begin{equation} 
g\prt{\boldsymbol{x}} = u_{max} - u\prt{\boldsymbol{x}}.
\label{eq_struc_frame_limit_state_function}
\end{equation}

We consider a maximum allowed displacement $u_{max}=50$ mm. The top-floor displacement of the building is computed using a finite element model, where geometrical and physical non-linearities are not considered. Additionally, the model takes 21 random variables as input. $P_{1}, P_2, P_3$ are the horizontal loads (see \figref{fig_SF_sketch}). The geometric and material properties, such as Young's moduli, moments of inertia, and cross-sections, are provided respectively $E_{4,5}$, $I_{6,\ldots, 13}$ and $A_{14,\ldots, 21}$, and their marginal distributions are given in \tabref{tab_SF_marginal}. \tabref{tab_SF_propoerties} displays the structural properties of each structural element depicted in \figref{fig_SF_sketch}. The dependence between the random input variables is modelled using a Gaussian copula, and its correlation matrix $\ma{R}$ is defined as follows:
\begin{itemize}
\item The correlation between the cross-sectional area of a given element and its inertia moment is set equal to $R_{A_i, I_i} = 0.95$;

\item The correlation between properties of different elements is set equal to $R_{A_i, I_j} = R_{I_i, I_j} =   R_{A_i, A_j}= 0.13$;

\item The correlation between Young's moduli is set equal to $ R_{A_i, A_j}= 0.90$;

\item The remaining correlations coefficients of $\ma{R}$ are set equal to 0.
\end{itemize}

\begin{table}[H]
    \centering
\begin{tabular}{llcc}
\hline Variable & Distribution & Mean & Standard Deviation \\
\hline$P_1(\mathrm{kN})$ & Lognormal & 133.454 & 40.04 \\
\hline$P_2(\mathrm{kN})$ & Lognormal & 88.97 & 35.59 \\
\hline$P_3(\mathrm{kN})$ & Lognormal & 71.175 & 28.47 \\
\hline$E_4\left(\mathrm{kN} / \mathrm{m}^2\right)$ & Truncated Gaussian & $2.1738 \times 10^7$ & $3.8304 \times 10^6$ \\
\hline$E_5\left(\mathrm{kN} / \mathrm{m}^2\right)$ & Truncated Gaussian & $2.3796 \times 10^7$ & $3.8304 \times 10^6$ \\
\hline$I_6\left(\mathrm{~m}^4\right)$ & Truncated Gaussian & $8.1344 \times 10^{-3}$ & $1.0834 \times 10^{-3}$ \\
\hline$I_7\left(\mathrm{~m}^4\right)$ & Truncated Gaussian & $1.1509 \times 10^{-2}$ & $1.2980 \times 10^{-3}$ \\
\hline$I_8\left(\mathrm{~m}^4\right)$ & Truncated Gaussian & $2.1375 \times 10^{-2}$ & $2.5961 \times 10^{-3}$ \\
\hline$I_9\left(\mathrm{~m}^4\right)$ & Truncated Gaussian & $2.5961 \times 10^{-2}$ & $3.0288 \times 10^{-3}$ \\
\hline$I_{10}\left(\mathrm{~m}^4\right)$ & Truncated Gaussian & $1.0812 \times 10^{-2}$ & $2.5961 \times 10^{-3}$ \\
\hline$I_{11}\left(\mathrm{~m}^4\right)$ & Truncated Gaussian & $1.4105 \times 10^{-2}$ & $3.4615 \times 10^{-3}$ \\
\hline$I_{12}\left(\mathrm{~m}^4\right)$ & Truncated Gaussian & $2.3279 \times 10^{-2}$ & $5.6249 \times 10^{-3}$ \\
\hline$I_{13}\left(\mathrm{~m}^4\right)$ & Truncated Gaussian & $2.5961 \times 10^{-2}$ & $6.4902 \times 10^{-3}$ \\
\hline$A_{14}\left(\mathrm{~m}^2\right)$ & Truncated Gaussian & $3.1256 \times 10^{-1}$ & $5.5815 \times 10^{-2}$ \\
\hline$A_{15}\left(\mathrm{~m}^2\right)$ & Truncated Gaussian & $3.7210 \times 10^{-1}$ & $7.4420 \times 10^{-2}$ \\
\hline$A_{16}\left(\mathrm{~m}^2\right)$ & Truncated Gaussian & $5.0606 \times 10^{-1}$ & $9.3025 \times 10^{-2}$ \\
\hline$A_{17}\left(\mathrm{~m}^2\right)$ & Truncated Gaussian & $5.5815 \times 10^{-1}$ & $1.1163 \times 10^{-1}$ \\
\hline$A_{18}\left(\mathrm{~m}^2\right)$ & Truncated Gaussian & $2.5302 \times 10^{-1}$ & $9.3025 \times 10^{-2}$ \\
\hline$A_{19}\left(\mathrm{~m}^2\right)$ & Truncated Gaussian & $2.9117 \times 10^{-1}$ & $1.0232 \times 10^{-1}$ \\
\hline$A_{20}\left(\mathrm{~m}^2\right)$ & Truncated Gaussian & $3.7303 \times 10^{-1}$ & $1.2093 \times 10^{-1}$ \\
\hline$A_{21}\left(\mathrm{~m}^2\right)$ & Truncated Gaussian & $4.1860 \times 10^{-1}$ & $1.9537 \times 10^{-1}$ \\
\hline
\end{tabular}
    \caption{Marginal distributions associated with the probabilistic input of the structural frame shown in \figref{fig_SF_sketch}. All Gaussian are truncated in the domain $[0,+\infty]$. The quoted moments refer to the full, untruncated Gaussian distributions.}
    \label{tab_SF_marginal}
\end{table}

\begin{table}[H]
    \centering
\begin{tabular}{cccc}
\hline Element & Young's modulus & Moment of inertia & Cross-sectional area \\
\hline$B_1$ & $E_4$ & $I_{10}$ & $A_{18}$ \\
$B_2$ & $E_4$ & $I_{11}$ & $A_{19}$ \\
$B_3$ & $E_4$ & $I_{12}$ & $A_{20}$ \\
$B_4$ & $E_4$ & $I_{13}$ & $A_{21}$ \\
$C_1$ & $E_5$ & $I_6$ & $A_{14}$ \\
$C_2$ & $E_5$ & $I_7$ & $A_{15}$ \\
$C_3$ & $E_5$ & $I_8$ & $A_{16}$ \\
$C_4$ & $E_5$ & $I_9$ & $A_{17}$ \\
\hline
\end{tabular}
    \caption{Material and geometrical properties of the structural elements depicted in \figref{fig_SF_sketch}.}
    \label{tab_SF_propoerties}
\end{table}

The associated noise-free probability of failure was computed via Monte Carlo simulation using $2\cdot  10^6$ samples and is equal to $5.07 \cdot 10^{-3}$. Its associated coefficient of variation is smaller than 1\%.

Similar to the previous examples, the model is corrupted by i.i.d Gaussian random variables $\cn\prt{0, \sigma_{\varepsilon}}$. In this case, however, it is possible to attribute a physical meaning to $\sigma_{\varepsilon}$ by considering a hypothetical scenario where measuring the top-floor displacement is possible. Under these conditions, the noise level can be associated with the accuracy of the measuring device. The accuracy varies according to the device and the techniques on which they rely. In the case of top-floor displacement, the measurement is usually done by GPS or vision-based techniques. Regarding GPS techniques, \cite{Nickitopoulou_2006, Lee_2006} report up to $\pm 10$ mm of accuracy, and some more recent papers  \citep{Sofi_2022} report a precision of a few millimeters. Regarding vision-based techniques, \cite{Guo_2020} report accuracies spanning from a few millimeters to a few centimeters, depending on the equipment and the distance of the measured object. Aiming to account for the significant variability in accuracy between different methods, we consider a noise level $\sigma_{\varepsilon}$ of $10$ mm, which results in a challenging denoising problem, especially when compared to the critical threshold of $50$ mm.

For this example, the initial ED comprises $N_{\textrm{ini}} = 150$ points sampled using LHS. This larger computational budget aims to address both the high dimensionality of the model and the presence of significant noise. The surrogate model is iteratively enriched based on the $U_N$ learning function, and similarly to the previous experiments, different enrichment strategies are tested, \emph{i.e.}, introducing 1, 3, and 10 points at a time. The maximum budget allowed for the enrichment step for this example is $1{,}000$ samples. Once again, 50 analyses are carried out.

\begin{figure}[H]
     \centering
     \begin{subfigure}[c]{0.49\textwidth}
         \centering
         \includegraphics[width=\linewidth]{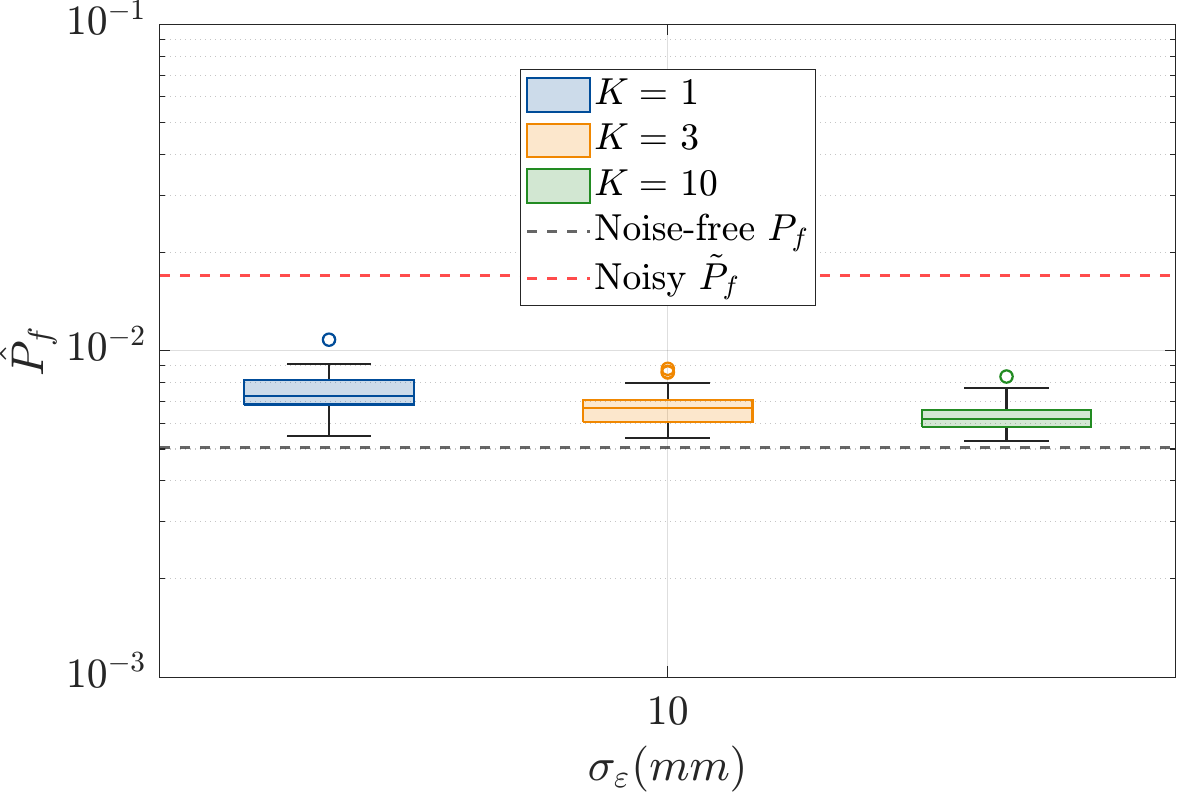}
         \caption{}
    \label{fig_SF_boxplots}
     \end{subfigure}
     \begin{subfigure}[c]{0.49\textwidth}
         \centering
         \includegraphics[width=\linewidth]{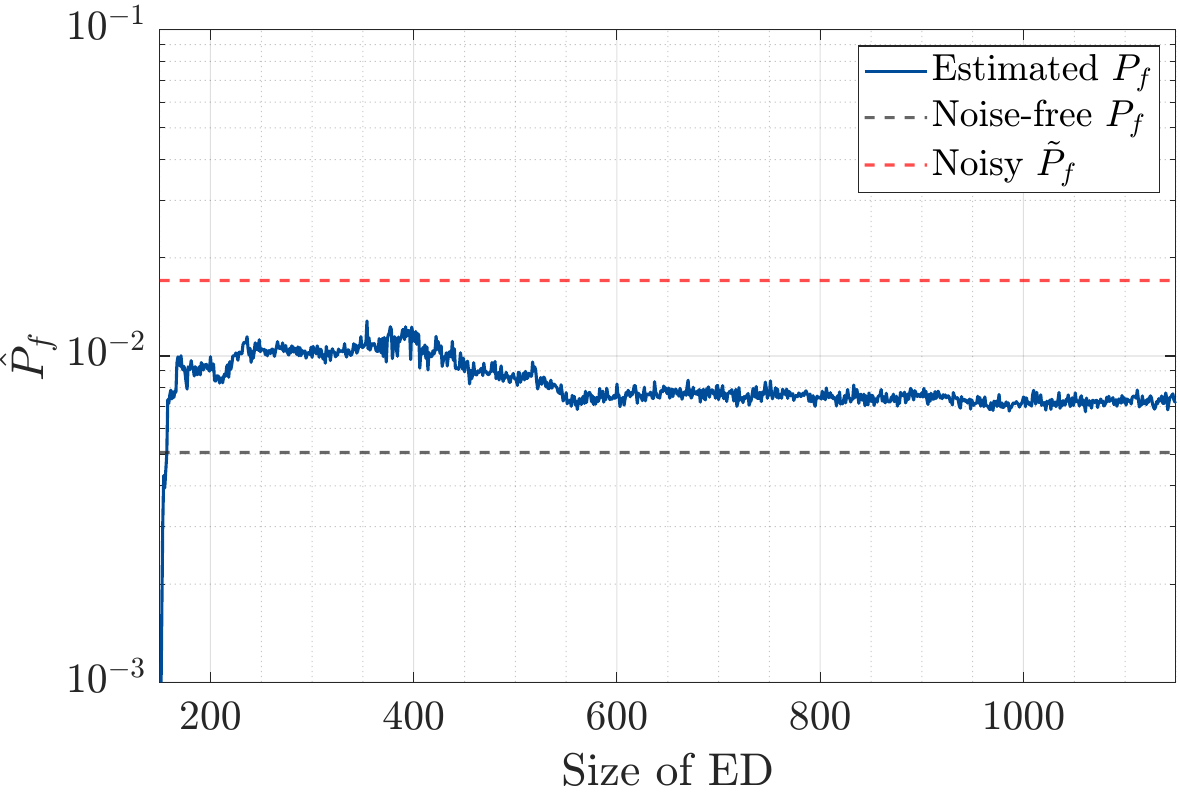}
         \caption{}
    \label{fig_SF_conv_curve}
     \end{subfigure}
     \caption{Structural frame -- (a): final probability of failure estimated using GPR on the limit-state function depicted in \eqrefe{eq_struc_frame_limit_state_function}, corrupted with a noise level of $10$ mm. (b): convergence curve of the experiment that resulted in the median $\hat{P}_f$ when enriched with one point at a time. The reference noise-free $P_f$ and the noisy $\tilde{P}_f$ are shown by the black and red dashed lines, respectively.}
\end{figure}

\figref{fig_SF_boxplots} displays the performance of the algorithm. The noisy and noise-free probabilities of failure are computed via Monte Carlo simulation and are $1.70 \cdot 10^{-2}$ and $5.07 \cdot 10^{-3}$, respectively. In this case, the noisy probability of failure is more than three times larger than its noise-free counterpart, causing a change in the order of magnitude of the probability of failure. Our algorithm converges to a  median estimate of $\hat{P}_f = 7.26 \cdot 10^{-3}$ for $K=1$. Unlike the previous examples, these results converge to a higher probability of failure compared to the noise-free reference. Nevertheless, we consider these results satisfactory, given the challenging nature of this example, which is characterised by both high dimensionality and high noise level.

\figref{fig_SF_boxplots} depicts the results obtained by the different enrichment strategies, showing that introducing multiple points simultaneously leads to more effective denoising of the limit-state function. These results could suggest that the $\un$ learning function favors exploitative behavior for $K=1$. Indeed, despite extensive testing, the exact reason for such results could not be determined. Nonetheless, given that overly exploitative behavior is not desired in learning functions, alternative ones are currently being investigated. For instance, SUR \citep{Bect2012} has been shown to be promising but it is computationally more expensive. Finally, \figref{fig_SF_conv_curve} shows the convergence curve for the replication that led to the estimation of the median denoised probability of failure. It is possible to notice that the presence of noise intensifies the already noisy behavior of the convergence curves.

\section{Conclusions}
In this paper, we formalize a new category of problems in reliability analysis, characterized by a limit-state function corrupted by noise. Our analysis reveals that the latter causes simulation-based reliability methods to estimate an incorrect, larger probability of failure. In this scenario, the main objective is to estimate the probability of failure associated with the noise-free limit-state function, \emph{i.e.}, the noise-free probability of failure.

We show that computing the noise-free probability of failure can be achieved through denoising. For this purpose, we propose the use of regression-based surrogate models, and focus particularly on Gaussian process regression. We demonstrate the feasibility of the proposed approach with the $R-S$ problem by showing that one can successfully recover the noise-free probability of failure using GPR when the experimental design is sufficiently large. However, as a large experimental design can be computationally demanding, we introduce an active learning approach to minimize the computational cost associated with the problem. Within this framework, we adopt the $U_N$ learning function, which is noise-aware.

Furthermore, we discuss the significant impact of the gradient around the limit-state surface in the denoising process. We also propose a novel methodology to create coherent noisy problems from noise-free benchmarks. Our results, obtained from two benchmark functions and a realistic structural frame, confirm the efficiency of our active learning approach in estimating the noise-free probability of failure in complex scenarios such as series systems, flat-gradient limit-state surfaces, and high-dimensional models. The final example demonstrates the feasibility of reliability analysis on experimental systems where the limit-state is \textit{measured} by an appropriate device instead of \textit{computed} by a simulator. Moreover, we show that the $\un$ learning function not only allows the method to effectively converge to the noise-free probability of failure but also maintains efficiency in a multi-point enrichment strategy. Notably, in one of the case studies, the multi-point enrichment improves the estimation of the probability of failure, suggesting that the $\un$ function might be prone to exploitative loops. To overcome this issue, we are currently exploring alternative noise-aware learning functions and enrichment strategies.

While our study focuses on homoskedastic Gaussian noise, the unique requirement for our method is unbiased noise. However, the effectiveness of Gaussian process regression in handling heteroskedastic non-Gaussian noise remains to be studied. Finally, developing a suitable stopping criterion for our approach, crucial for optimizing its efficiency and applicability, is currently under investigation.

\appendix
\section{Limit behavior of the $U_N$ learning function}\label{appendix}
In this section, we show that, if the predicted Gaussian process variance $\sigma^2_{\hat{g}}\prt{\boldsymbol{x}}$ converges to the estimated noise level $\gpn^2$, the optimization problem described in \eqrefe{eq_optim_Un} becomes non-informative. With this aim, we first compute the one-step look-ahead variance $\sigma^2_{\hat{g}+1}(\boldsymbol{x})$ when $\sigma^2_{\hat{g}}\prt{\boldsymbol{x}} \rightarrow \gpn^2$, as follows:
\begin{equation}\label{eq:limsigma2g+1}
\lim_{\sigma^2_{\hat{g}}\prt{\boldsymbol{x}} \rightarrow \gpn^2}  
                  \sigma^{2}_{\hat{g}+1}(\boldsymbol{x})= \lim_{\sigma^2_{\hat{g}}\prt{\boldsymbol{x}} \rightarrow \gpn^2} \sigma^{2} _{\hat{g}}(\boldsymbol{x})\frac{\gpn^{2}}{\sigma^{2}_{\hat{g}}(\boldsymbol{x})+\gpn^{2}}= \frac{\gpn^2}{2}.
\end{equation}

Inserting \eqrefe{eq:limsigma2g+1} into \eqrefe{uq_Un_learning_function} and considering that $\mu_{\hat{g}+1}\prt{\boldsymbol{x}} = \mu_{\hat{g}}\prt{\boldsymbol{x}}$, we obtain:
\begin{equation}          
U_N\prt{\boldsymbol{x}} =\Phi\left(-\frac{\abs{\mu_{\hat{g}}(\boldsymbol{x})}}{\gpn}\right)-\Phi\left(-\frac{\sqrt{2}\abs{\mu_{\hat{g}}(\boldsymbol{x})}}{\gpn}\right).
\label{eq_Un_lim_noise}
\end{equation}

Then, we show that \eqrefe{eq_Un_lim_noise} has an analytical optimum. For the sake of simplicity, we introduce the following change of variables:
\begin{equation}
    t=\frac{\abs{\mu_{\hat{g}}(\boldsymbol{x})}}{\gpn}.
\end{equation}

\noindent The argument $t \geqslant 0$ that maximizes $U_N\prt{t}$ may be obtained by setting its derivative to zero and finding the critical points. The derivative of  $U_N\prt{t}$ with respect to $t$ reads:
\begin{equation}
\begin{split}
    \dfrac{\mathrm{d} \, U_N\prt{t}} {\mathrm{d}t}=    \dfrac{\mathrm{d}}{\mathrm{d}t} \prt{\Phi\left(-t\right)-\Phi\left(-\sqrt{2}t\right)}.
\end{split}
\end{equation}

\noindent Noting that $\Phi\prt{t}=\int_{-\infty}^t\frac{1}{\sqrt{2 \pi}} \exp\prt{\frac{-x^2}{2}} \textrm{d}x$ and by differentiating under the integral sign, the derivative of $\un\prt{t}$ vanishes for $t$ such that:
\begin{equation}\label{eq:gradUN}
    \exp\prt{-\frac{t^2}{2}} + \prt{-\sqrt{2}}\exp\prt{-t^2} =0
\end{equation}

Solving \eqrefe{eq:gradUN} yields the positive solution $t = 
 \sqrt{\ln{2}}$, which means that the $\un$ learning function reaches its global maxima when:
\begin{equation}
    {\abs{\mu_{\hat{g}}(\boldsymbol{x})}}={{\gpn}}\cdot\sqrt{\ln{2}}.
    \label{eq_global_maximum}
\end{equation}

From this result, we can conclude that if $\sigma^2_{\hat{g}}\prt{\boldsymbol{x}} \rightarrow \gpn^2$,  the learning function will select the points of the candidate set $\xc$ that best satisfy the condition shown in \eqrefe{eq_global_maximum}. Such points are however not close to the limit-state surface when $\sigma_n$ is large.

\section*{Acknowledgments}
The support of European Union’s Horizon 2020 research and innovation program under the Marie Skłodowska-Curie grant agreement No 955393 is greatly acknowledged.

\newpage
\bibliography{cas-ref}
\end{document}